# Search for emission-line galaxies towards nearby voids. Observational data*

**Cristina C. Popescu**[1,4], **Ulrich Hopp**[1,2], **Hans Jürgen Hagen**[3], **and Hans Elsässer**[1]

[1] Max Planck Institut für Astronomie, Königstuhl 17, D–69117 Heidelberg, Germany
[2] Universitätssternwarte München, Scheiner Str.1, D–81679 München, Germany
[3] Hamburger Sternwarte, Gojenbergsweg 112, D–21029 Hamburg, Germany
[4] The Astronomical Institute of the Romanian Academy, Str. Cuţitul de Argint 5, 75212, Bucharest, Romania

**Abstract.** We present the observational results of our search for emission-line galaxies (ELG) towards nearby voids. In order to find ELG, we started a survey using the IIIa-J objective prism plates from the Hamburg QSO Survey. The plates are digitized and an automatic procedure was applied to select the candidates. Digitized direct plates were used to determine coordinates and to reject overlaps between spectra. The accuracy of the coordinates is $\pm 2''$. A total area of $1248\,\mathrm{deg}^2$ was scanned, distributed in four different regions.

All the selected objects were observed with follow-up spectroscopy. We have obtained a final sample of 203 objects, of which 196 are emission-line galaxies, four are galaxies with absorption lines and three are QSOs. Almost half of our objects are newly discovered ones and three quarters of the given redshifts are new. Our sample contains mainly high ionization galaxies and is less sensitive in the detection of low-ionization objects.

The apparent magnitudes, as derived from the objective prism plates, range between $15.0 \leq B \leq 19.5$. The sample is dominated by nearby galaxies, with a peak in the redshift distribution at cz=4500 km/s.

## 1. Introduction

Recent redshift surveys of galaxies (e.g. the Center for Astrophysics Survey (CfA)), revealed that bright galaxies are distributed in sheet-like structures which surround large voids. This has been a subject of intense investigation since the proposal of such structures by Zeldovich et al. (1982). Still under debate is the question whether all galaxies follow this non-uniform distribution or if less luminous galaxies are more equally distributed. Given the limitation of the actual surveys, the observed emptiness of the voids may be a result of observational bias. From an observer's point of view, the galaxy maps may reflect special observational selection effects in surface brightness, integral magnitude or diameter. From the theoretical point of view, there has been the prediction (e.g. Kaiser 1986, Bardeen 1986) that luminous galaxies were formed preferentially in high-density regions of the Universe, thus giving us a biased view of the large-scale distribution of matter.

Several studies of the spatial distribution of galaxies, especially of dwarf galaxies, were carried out to overcome some of these biases (e.g. Binggeli et al. 1990, Thuan et al. 1991). Unfortunately the results were contradictory and no definitive conclusions were drawn. Some interesting results came from the study of the spatial distribution of emission-line galaxies (ELG) (Salzer 1989, Weistrop et al. 1992). These objects are intrinsically small, very compact and with low luminosities. In some cases they are so compact that they look almost stellar, with no obvious underlying galaxy. Such galaxies can easily be missed by surveys that select their candidates on morphological criteria or discriminate stars from galaxies by the apparent diameter of the latter, and thus are good candidates to fill up the voids. HII galaxies can be recognised because of their typical emission-line spectrum and an efficient way to find them is to use objective prism plates.

We therefore started a project to search for ELG towards nearby voids, using an objective prism survey. This project belongs to a larger one, that has the aim of finding faint galaxies within the voids (Hopp and Kuhn 1995, Hopp et al. 1995). Previous results on this project showed that while the central part of the voids remains free of galaxies, some of the dwarfs populate the outskirts of the voids. Most of these dwarfs are emission-line galaxies.

During the last two decades much effort has been spent to discover large samples of ELG. One method is to use multiple exposures on the same plate through different filters and search for UV excess objects, as was done by the Kiso UV-Excess Galaxy Survey (Takase and Miyauchi-Isobe 1988) and the Montreal Blue Galaxy Survey (Coziol et al. 1993, Coziol et al. 1994). A second method is to use low resolution objective prism Schmidt plates and search for galaxies with strong UV-continuum. This method was used by Markarian (1967) for the First Byurakan Spectral Sky Survey. Perhaps the most unambiguous method is to use objective prism surveys and search directly for the presense of emission-lines.

*Send offprint requests to*: Cristina C. Popescu
*Partly based on observations obtained at the European Southern Observatory, La Silla, Chile

Haro (1956) was the first to show that moderate-dispersion plates can show emission-lines of [OII], [OIII], the Hydrogen Balmer series and other elements. The method was successfully applied by Smith (1975) for the Tololo Objective Prism Survey (see also Smith et. al 1976, Bohuski, Fairall and Weedman 1978). They used the 61-cm Curtis Schmidt telescope at Cerro Tololo to take IIIa-J (1740 Å/mm at $H_\beta$) plates. The same technique was then applied for the University of Michigan Survey (UM) (MacAlpine et al. 1977a,b,c, MacAlpine & Lewis 1978, MacAlpine & Williams 1981), the Calan-Tololo Survey (Maza et al. 1989), the Wasilewski Survey (Wasilewski 1983), and the Surace & Comte Survey (Surace 1993, Surace & Comte 1994). Sometimes a combination between emission-line and UV-excess criteria was applied, as in the Case Survey (Sanduleak & Pesch 1982, 1984, 1987, 1989, 1990, Pesch & Sanduleak 1983, 1986, 1988, 1989, Pesch et al. 1991, Stephenson & Pesch 1992, Pesch et al. 1995). An alternative to the IIIa-J plates is to use red IIIa-F objective plates to search for $H_\alpha$ emission-lines. These surveys seem to be complementary to the IIIa-J surveys, as discussed by Zamorano et al. 1994. The main $H_\alpha$ surveys were: Wamsteker et al. 1985, Moody et al. 1987, and Universidad Complutense de Madrid Survey (UCM) (Zamorano et al. 1990, 1994, Gallego 1995). Smaller areas of sky were searched by Kunth et al. 1981 (two IIIa-F plates) and Kunth & Sargent 1986 (three IIIa-J plates in the Sculptor Region). A combination between both IIIa-J and IIIa-F plates was used by Kinman (1984). Finally, the Second Byurakan Spectral Sky Survey (SBSS)(Markarian et al. 1983a,b, 1984, Stepanian et. al 1990, 1991), which is in progress, observed each field in three different colours: UV and blue region (IIIa-J plates), green region (IIIa-J + GG495) and red region (IIIa-F plates + RG2).

Despite the long lists of candidates that were produced, most of these surveys did not undertake systematic and complete follow-up spectroscopy of their candidates. The most complete and deep sample of ELG remains probably the Michigan Survey. For UM List IV and List V, a sample with complete follow-up observations was obtained (Salzer et. al 1989). A similar sample was obtained for List I and List II of the Case Survey (Salzer et al. 1995). On the other side, most of these surveys searched by eye their candidates and thus cannot provide objective selected samples. Modern techniques for the digitization of Schmidt plates, as pioneered in Cambridge (Kibblewhite et al. 1984; Irwin & Trimble 1984) or Edinburgh (Cooke et al. 1986), now offer the possibility of performing the object searches using controlled selection criteria. These techniques were already used for QSO searches (Clowes et al. 1984; Hewett et al. 1985, Wisotzki 1994, Hagen et al. 1995). The Surace & Comte Survey is the first survey for ELG that selected their candidates using digitized plates. Details of the digitized procedure are given in Surace & Comte 1994 (see also Surace 1995). The UCM survey applied a digitized procedure in order to test their eye selected sample (Alonso 1995).

A new era in the survey techniques was opened by the use of grisms and CCD cameras as detectors (e.g. Palomar Transit Grism Survey, Schneider et al. 1994). While these kind of surveys have the advantage of going deeper, their small field of view makes the photographic plate surveys still competitive when covering wide fields.

In this paper we present an objective prism survey that used digitized plates and an automatic procedure to search for ELG candidates. All the objects selected from the plates were observed with follow-up spectroscopy, thus providing a complete sample of emission-line galaxies. The paper describes the selection of the sample and the follow-up observations. The study of the spatial distribution of our sample of emission-line galaxies, as well as their properties, will be given elsewhere.

The paper is organized as follows. In § 2 we describe the method used to select the candidates from the objective plates. In § 3 the follow-up spectroscopy is presented, in § 4 we discuss the results of our survey and § 5 contains the conclusions.

## 2. Selection of candidates on objective prism plates

We used published cone diagrams to identify nearby ($v_R \leq 7500 km/s$) voids. Four void regions were selected according to the following criteria: diameter larger than 20 Mpc ($H_0 = 75\, km/s/Mpc$), completely empty of CfA galaxies, galactic latitude $b \geq 30°$. The location of the four void regions can be found in Table 1, where we give the exact coordinates of all the fields surveyed.

The ELG candidates were selected on the objective prism plates taken in the frame of the Hamburg Quasar Survey (HQS) (Hagen et al. 1995). The plates are taken with the 80-cm Schmidt-telescope at the German-Spanish Observatory, Calar Alto (Spain). The telescope is equipped with a 1.7° objective prism producing a dispersion of 1390 Å/mm at $H_\gamma$. The 24 cm× 24 cm hypersensitized Kodak IIIa-J plates (spectral range: 3400 Å-5500 Å), covering a field of 5.5° × 5.5°, were scanned with low resolution (LR), using a PDS 1010G microdensitometer. The LR mode used a 100 $\mu$m slit running perpendicular to the direction of dispersion. After on-line background reduction and object recognition, LR spectra were stored on optical disks. Automated search software was applied to the LR digitized data to select spectra in a specific parameter space. The two parameters used were the density sum of intensity ("brightness") of the integrated spectra and the slope of the continuum ("colour"), which is again a function of the density sum. Brighter objects are redder because the sensitivity of the photographic plate increases with wavelength. The selected spectra were re-scanned individually with high resolution (HR) (30 $\mu$m slit) and the final digitized spectra were visually inspected for emission lines.

The range in density sum excludes very bright objects, that are not needed for our study, but also very faint objects, that lie very close to the detection limit of the plate. The limit at the very bright end helps us to decrease the number of uninteresting candidates, while bright galaxies (brighter than our threshold) with emission are still found due to their HII regions. These knots are found as individual objects and their brightnesses are therefore fainter than those of the underlying galaxies. Nevertheless, these bright galaxies are already known and contained in galaxy catalogs like the CfA. The incompleteness due to the faint limit in the low-density region was analyzed as follows. For three plates we extended our search to the very faint end (which is only 20% of the whole range used), and we found twice (or even more) as many candidates as in the normal range. Most of these extra spectra were very noisy and could not be classified. Thus the efficiency of finding interesting objects was rather reduced and the number of objects and the quality of the spectra depended more strongly on plate quality. Our tests showed that we can miss some faint emission-line galaxies, on the order of 2 per plate (0.06 $deg^{-2}$), but one can correct for these numbers. Due to our automated search software, the incompleteness at the very faint end of the "brightness" parameter can be quantified. The other parameter, the slope of the continuum, is set so that only the bluest objects are selected. To do this we estimate the slope as a function of density sum, and for each bin of density sum, a fraction *f=0.15* of the bluest objects are considered. The colour selection causes our sample to be dominated by very blue galaxies. In order to test this selection effect, we searched two plates without any restriction in the colour parameter. Our tests showed again that we can miss some emission-line galaxies that are redder than our selection criteria. These galaxies are usually strong emission-line objects and have also strong continua, that is produced by an underlying older population of stars. As we are more interested in dwarf emission line galaxies, we believe that our blue criteria help us to select a sample that is less contaminated by other kinds of emission-line galaxies. We will refer to our selection criteria as the blue criteria, though it includes a selection in a two parameter space.

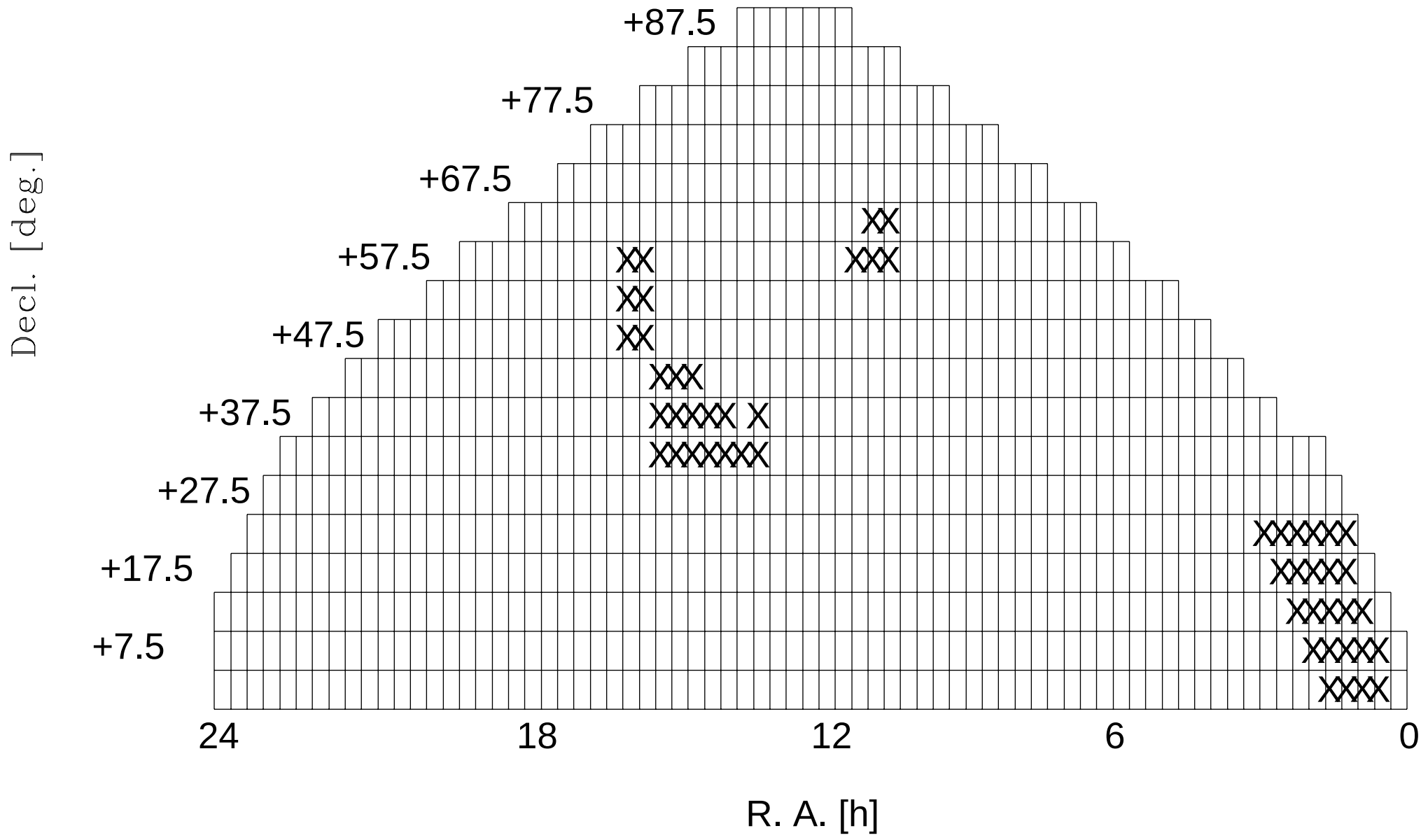

**Fig. 1.** The map of the Northern Sky, divided into fields of $5.0 \times 5.0\,deg^2$. We marked with crosses the plates that were inspected in our ELG Survey. Each plate covers a field of $5.5 \times 5.5\,deg^2$, thus allowing an overlap between neighbour plates of $0.5°$. Field centres are provided in Table 1.

The number of candidates for high resolution varies between 1500 and 3500 per plate, depending on the plate quality and on the galactic latitude. Those HR spectra which show strong [OIII]$\lambda$5007 lines are selected the final sample of candidates. The [OIII] line appears as a distinctive peak near the green head of the objective prism spectra. Often the [OII]$\lambda$3727 doublet can be identified too. Some typical HR objective prism spectra are given in the left panels of Fig. 4. The aim of the Hamburg Survey is to cover every Schmidt field with two plates, in order to allow the distinction between real emission-lines and plate defects. Unfortunately the regions we scanned were not always covered by two plates, as the Hamburg sur-vey is still in progress. For the fields where we had to rely on only one plate, the success rate of our selection (see below) is lower because of the addition of a few false candidates that did not show any emission lines in the follow-up observations. Also there were not always high quality plates available for every field. The final sample contains 234 candidates, distributed in four different regions, in a total scanned area of 1248 $\deg^2$. With a surface density of 0.19 candidates/$\deg^2$, we will refer to this sample as the first priority sample. In Fig. 1 we show the map of the scanned fields. In Table 1 we give the coordinates of the centre of the corresponding plates, separate for every void region. One plate covers a field of $5.5° \times 5.5°$ and there is an overlap between neighbour plates of $0.5°$.

**Table 1.** The coordinates of the centre of the plates that were scanned in our ELG Survey. The first column gives the void region (from 1 to 4) for which the plates were selected

| | |
|---|---|
| 1 | $\delta = 2.5°$: $\alpha = 00^h20^m$; $00^h40^m$; $01^h00^m$; $01^h20^m$; |
| | $\delta = 7.5°$: $\alpha = 00^h20^m$; $00^h40^m$; $01^h00^m$; $01^h20^m$; $01^h40^m$; |
| | $\delta = 12.5°$: $\alpha = 00^h20^m$; $00^h41^m$; $01^h01^m$; $01^h21^m$; $01^h41^m$; |
| | $\delta = 17.5°$: $\alpha = 00^h21^m$; $00^h42^m$; $01^h03^m$; $01^h23^m$; $01^h44^m$; |
| | $\delta = 22.5°$: $\alpha = 00^h00^m$; $00^h21^m$; $00^h43^m$; $01^h04^m$; $01^h26^m$; |
| 2 | $\delta = 57.5°$: $\alpha = 08^h12^m$; $08^h47^m$; $09^h22^m$; |
| | $\delta = 62.5°$: $\alpha = 08^h00^m$; $08^h40^m$; |
| 3 | $\delta = 32.5°$: $\alpha = 13^h10^m$; $13^h33^m$; $13^h56^m$; $14^h19^m$; $14^h43^m$; |
| | $\delta = 37.5°$: $\alpha = 12^h37^m$; $13^h25^m$; $13^h50^m$; $14^h14^m$; $14^h39^m$; $15^h03^m$; |
| | $\delta = 42.5°$: $\alpha = 14^h24^m$; $14^h50^m$; $15^h16^m$; |
| 4 | $\delta = 47.5°$: $\alpha = 16^h00^m$; $16^h28^m$; |
| | $\delta = 52.5°$: $\alpha = 16^h42^m$; $17^h13^m$; |
| | $\delta = 57.5°$: $\alpha = 16^h59^m$; $17^h34^m$; |

Though our main selection criteria was the presence of emission features, we have also second priority candidates that were chosen because of their very blue continuum. With a surface density of about 1 candidate/$\deg^2$, these objects do not show any reliable emission features and for this reason they were considered as a separate sample.

Digitized direct plates were used to determine coordinates, to reject artificially blue objects created by overlaps of two spectra, and finally to produce finding charts. For details of the digitization procedure and astrometric determinations see Hagen et al. (1995).

## 3. Follow-up spectroscopy

We observed our candidates during several observing runs, most of them being carried out with the 2.2 m telescope at the German-Spanish Observatory at Calar Alto (Almeria, Spain). Table 2 lists the dates of the different runs, as well as the instruments and particularities of the set-up used each time.

At the 2.2 m telescope in Calar Alto we used a Boller & Chivens standard spectrograph (runs: 2,4,7) and also the new CAFOS (Calar Alto Faint Object Spectrograph) focal reducer system and grisms (run 3). A few observations were made with the Prime Focus Focal Reducer at the Calar Alto 3.5 m telescope. Runs 1 and 6 were not dedicated to our project, therefore our candidates (only 8) were observed as backup objects. During all the Calar Alto observations, a Tektronix 1024 by 1024 24$\mu$ pixel size CCD chip served as the detector while a coated Thompson 1024 by 1024 19$\mu$ chip was used at La Silla (run 6). In all cases, a long slit of about 2 arcsec width was used. With the focal reducers, we always obtained a blue and/or a red image which served for the measurements of magnitudes and to estimate morphological types. The usual calibration measurements (bias-and flat-frames for the CCD correction, wavelength calibration frames and flux standard stars) were carried out.

Typical exposure times were 10-15 minutes, depending primarily on the strength of the emission-lines: strong-line objects were not observed as long as weaker line galaxies. The exposure time depends also on the brightness of the galaxy, though not so strongly. For example, for very faint objects with almost no underlying continuum but with strong emission features, an integration of 10 minutes brought enough fluxes in the emission lines to properly reduce the spectrum. By contrast, brighter objects, with strong continuum but very weak lines (or no obvious emission features in the objective spectra) had to be integrated longer.

The frames were biased and flatfield-corrected. For the extraction of the 1-dimensional spectra from the 2-dimensional data, the optimal extraction algorithm of Horne (1986) was used. The spectra were rebinned to a linear wavelength scale using a third or fifth order polynomial fitted to the dispersion curve of the comparison spectra. A flux calibration was applied, and finally the wavelength scale was checked by comparison with the night-sky lines. A more detailed description of the reduction procedure is given by Stickel et al. (1993).

Once fully reduced, the emission lines in each spectrum were measured by fitting a Gaussian. The quoted redshifts were derived as means of the redshifts determined from the individual strong lines, and the errors of the redshifts were calculated as error of the mean. The observed redshifts were further corrected for the motion of the Earth and transformed in heliocentric redshifts. The calculated errors were compared with the deviations obtained from the night-sky line measurements. We estimate that for the spectra taken in the runs 3,4,7, which consist of 85% of our data, the errors are $\Delta$z=0.0001. This value can increase up to $\Delta$z=0.0002 for some of the galaxies with very noisy spectra. For the rest of our spectra (15%) the errors are bigger, up to $\Delta$z=0.0005. During run 7, some planetary nebulae with known redshifts were observed (NGC6543, NGC0040, NGC1501, NGC2022, NGC2392, PN212.0), in order to check the quality of our wavelength calibration. The reference values of the planetaries were taken from The Strasbourg-ESO Catalogue of Galactic Planetary Nebulae. (Acker et al. 1992)

In Fig. 2 we plotted the observed velocities versus the reference velocities and applied a linear fit. As one can see, the differences between our velocities and those from literature are in the range of the assumed errors. We obtain a correlation

**Table 2.** Dates and set-up of the spectroscopic observations: BCCS: Boller & Chivens Cassegrain Spectrograph, CAFOS 2.2 : Calar Alto Faint Object Spectrograph at the 2.2 m telescope, PFFR: Prime focus focal reducer, EFOSC2: ESO Faint Object Spectrograph and Camera 2, CA: Calar Alto, LS: La Silla

| No. | Dates | Telescope | Instrument | Spectral Range (Å) | Spectral Resolution (Å) |
|---|---|---|---|---|---|
| 1 | 1994 Jan 2-3 | 2.2 m CA | BCCS | 3600-6500 | 6.5 |
| | | | | 6200-9000 | 6.5 |
| 2 | 1994 Jan 4-9 | 2.2 m CA | BCCS | 3700-9000 | 17 |
| 3 | 1994 June 8-11 | 2.2 m CA | CAFOS 2.2 | 3900-8200 | 18 |
| 4 | 1994 Oct 9-12 | 2.2 m CA | BCCS | 3600-9000 | 12 |
| 5 | 1994 Oct 13 | 3.5 m CA | PFFR | 3500-7600 | 20 |
| 6 | 1994 Dec 10 | 2.2 m LS | EFOSC2 | 3500-5500 | 12 |
| 7 | 1995 Jan 31-Feb 3 | 2.2 m CA | BCCS | 3460-8600 | 12 |

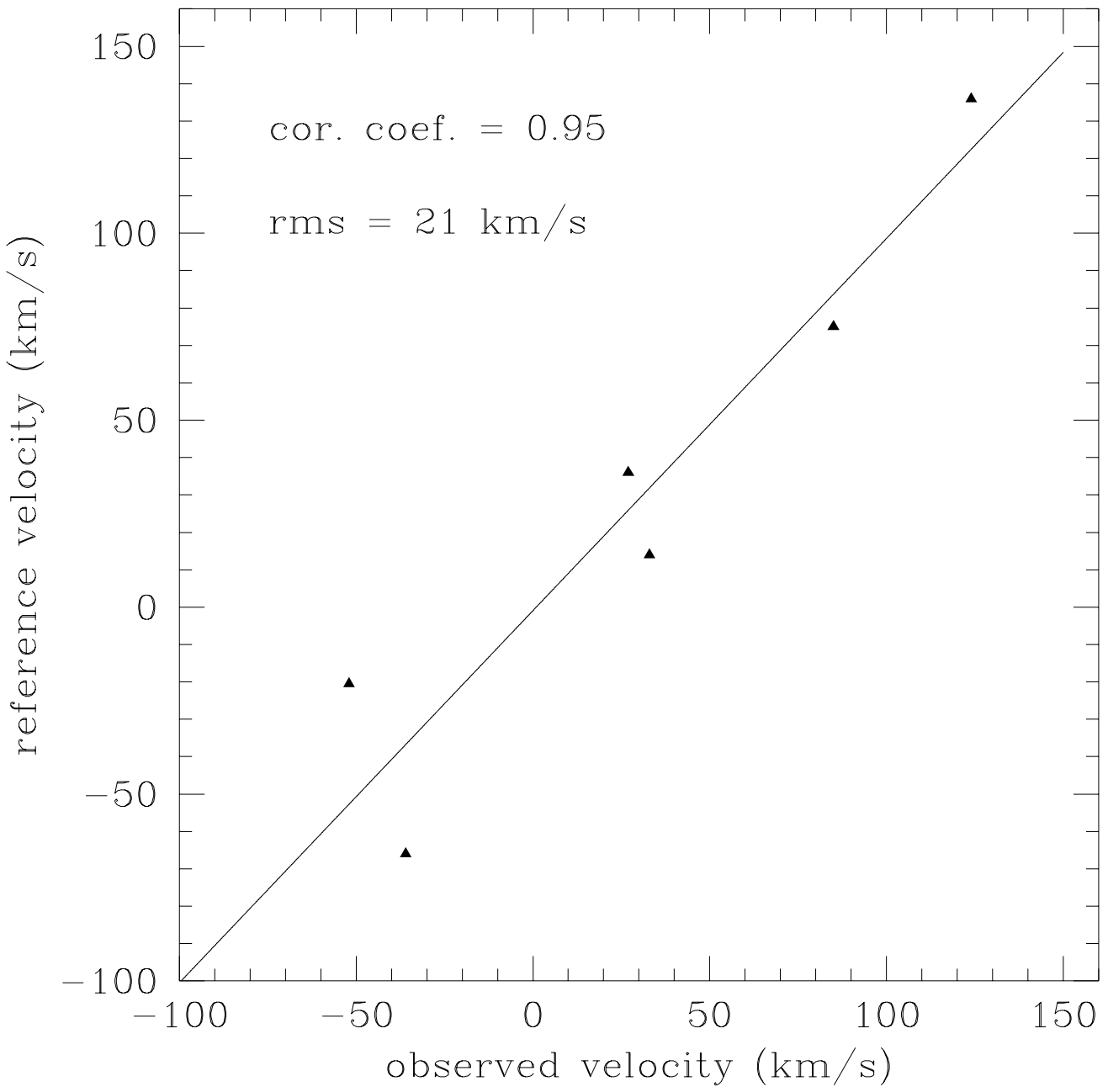

**Fig. 2.** The observed velocities of the planetary nebulae used to check the redshift errors, versus their reference velocities. A linear fit is applied to the data points and a correlation coefficient of 0.95 is obtained.

coefficient of 0.95 and a rms of 21 km/s, in agreement with our expectations.

A further test came from the observations of some candidates that had redshifts given in the literature. In Table 3 we give the names of the galaxies observed together with their velocities, both measured and from the literature. In the last column we indicate the difference between our values and the reference ones. These galaxies were measured during the runs for which we estimated an error of 0.0001 in z. Besides UM 306, for which we obtained a big difference, the rest show differences of about 0.0001. In order to see if the estimated error of 0.0001 in $z$ is constant through all the runs where we assumed such an error, we observed a few candidates twice, in different runs. The results of these tests indicate that the errors do not change with the run.

**Table 3.** Candidates with available redshifts from literature, observed with the purpose of testing the redshift accuracy.

| name | our redshift | reference redshift | difference in redshift |
|---|---|---|---|
| CG 0419 | 0.0114 | 0.0114 | 0.0000 |
| DDO 166 | 0.0031 | 0.0032 | 0.0001 |
| UM 065 | 0.0210 | 0.0210 | 0.0000 |
| DDO 13 | 0.0021 | 0.0021 | 0.0000 |
| UM 306 | 0.0170 | 0.0164 | 0.0006 |

Before starting the observations, we checked our lists of candidates with the present catalogs of galaxies and emission-line objects, in particular with the NASA Extragalactic Data Base (NED). Candidates found by other surveys with available redshifts were normally not reobserved. There were cases in which the candidates were contained also in other catalogs, but with unknown redshifts. We observed these, together with the newly discovered objects. All first priority candidates with unknown redshifts (182 out of 234) were measured and also some of our second priority ones (83). The latter were observed to test their nature and we did not intend to be complete with these objects. From the observed 182 first priority candidates, 126 showed emission lines, the rest being failures (either featureless spectra or stars). If we consider the whole sample of 234 first priority candidates (32 were objects found already in the literature with available redshifts), the success rate of our selection is 76%. Some of our failures come from the fields covered only by one plate (see § 2). But more important is the dependence of the success rate with the apparent magnitude (see p. 290, Hopp et al. 1995). For brighter objects the success rate is almost 100%, decreasing for the fainter ones.

The observed objects are listed in Table 4, which contains only the objects for which a redshift was assigned (failures not included). Most of our spectra show the emission lines of [OII]$\lambda\lambda$3727, H$\beta\lambda\lambda$4861, [OIII]$\lambda\lambda$4959, [OIII]$\lambda\lambda$5007, H$\alpha\lambda\lambda$6563 and [SII](blend)$\lambda\lambda$6724. Many spectra also show the fainter emission lines of ion species like [NeIII], [HeII], [OI], [OII], [AIII], as well as some further Balmer lines, from H$\gamma$ to H$\zeta$. These are mainly high ionization type spectra, which dominate our sample of ELG. We have even cases with very strong [OIII]$\lambda\lambda$4363, despite the short exposures that were used to take the spectra. Nevertheless there are also intermediate ionization spectra as well as low ionization ones, the latter coming mainly from the second priority sample. A few cases contain objects in which only H$\alpha$ and [SII] were detected. The spectra with only one line detection (four cases) are marked with an asterisk in Table 4. The table is organized as follow:

- **Column (1)** gives the name of our objects, which is built with the prefix HS (from Hamburg Survey), followed by the first four digits of the 1950.0 right ascension and declination;
- **Column (2)** gives alternate designations for those galaxies contained in other catalogs. The abbreviations are explained in the List of Abbreviations, given at the end of the table.
- **Column (3)** and **(4)** give the 1950.0 positions. The coordinates are derived from the Hamburg direct plates and have an accuracy of $\pm 2''$. There are few cases in which some HII regions in normal galaxies were found. The coordinates refer then to the position of the HII region rather than to the centre of the galaxy. Special remarks are made for each case.
- **Column (5)** gives the heliocentric redshifts,
- **Column (6)** the B magnitudes as derived from the Hamburg objective prism plates ($B_H$) and
- **Column (7)** the magnitudes from literature, $B_L$ (where available). For details of the magnitude calibration see Engels et al. (1994). The magnitude accuracy is $0.5^m$. Sometimes our magnitudes were fainter then those found in literature, because they refer only to the emission region, and not always to the underlying galaxy.
- **Column (8)** contains a flag S (from selection) that gives the selection criteria used to select the candidates from the objective prism spectra: E for emission candidate (first priority) and B for blue candidate (second priority) (see § 2 ).
- **Column (9)** contains either the number of the run in which the object was observed (with the prefix o - from observed) or the code of the reference, when the object had available redshift (with the prefix l - from literature). The numbers of the observing runs are the same as listed in Table 2. The codes for references are explained at the end of the table, in the "References to Table 4".
- **Column (10)** contains special remarks.

Finding charts of all our newly discovered objects can be found in Fig. 3. We also give the finding charts of some of our objects that were previously known as IRAS sources but without any follow-up spectroscopy. The finding charts were prepared by means of the Palomar Sky Survey plates, digitized and distributed on CD-ROM by the Space Telescope Science Institute. Fields are $10' \times 10'$. North is up, and east is to the left.

Some typical examples of our slit spectra are given in the right panels of Fig. 4, while the corresponding HR objective prism spectra can be found in the left panels. In the last slit spectrum, HS0153+2205, the [OIII] line is under the limit of a clear detection, the only lines being the H$\alpha$+NII. The corresponding objective prism spectrum was very noisy, and what appeared to be an [OIII] line was in fact spurious. We also give the spectrum of a Sy 1 galaxy, HS0814+6439. The identified emission lines are listed in Table 5 while the lengthy discussion of their strength is left for a separate paper. Those objects which are certainly stars are listed in Table 6.

Table 1

| (1) object name | (2) other names | (3) R.A. | (4) Decl. | (5) redshift | (6) $B_H$ | (7) $B_L$ | (8) S | (9) run or reference | (10) remarks |
|---|---|---|---|---|---|---|---|---|---|
| HS0000+2252 | IRAS F00000+2252 | 00 00 01.0 | +22 52 20 | 0.0730 | 16.3 | 17.33 | E | o4 | IrS |
| | NPM1G +22.0001 | | | | | | | | |
| HS0000+2422 | NPM1G +21.0002 | 00 00 20.9 | +24 22 35 | 0.0382 | 17.0 | | E | o4 | IrS |
| HS0006+2133 | IRAS 00067+2133 | 00 06 43.7 | +21 33 09 | 0.0774 | 17.5 | 16.9 | E | l12 | IrS |
| | IRAS F00067+2133 | | | | | | | | |
| | [dKM92] 003 | | | | | | | | |
| HS0013+0809 | | 00 13 02.0 | +08 09 37 | 0.0844 | 19.3 | | E | o7 | |
| HS0013+1942 | | 00 13 14.4 | +19 42 08 | 0.0258 | 17.1 | | E | o4 | abs |
| HS0013+2241 | | 00 13 15.0 | +22 41 45 | 0.0217 | 17.7 | | E | o4 | |
| HS0016+1449 | | 00 16 35.7 | +14 49 44 | 0.0147 | 17.6 | | E | o7 | |
| HS0020+0656 | UM 029 | 00 20 18.4 | +06 56 53 | 0.051 | 18.0 | | E | l77 | |
| HS0021+1347 | | 00 21 50.5 | +13 47 33 | 0.0144 | 16.5 | | E | o2 | |
| HS0022+0014 | UM 241 | 00 22 46.1 | +00 14 54 | 0.0139 | 16.3 | | E | o4 | |
| HS0022+1945 | | 00 22 44.8 | +19 45 23 | 0.0926 | 18.2 | | E | o5 | |
| HS0024+2314 | | 00 24 15.4 | +23 14 37 | 0.0238 | 18.0 | | E | o4 | |
| HS0024+1022 | | 00 24 32.2 | +10 22 28 | 0.0071 | 17.0 | | E | o2 | |
| HS0026+0332 | | 00 26 52.4 | +03 32 49 | 0.0410 | 17.5 | | E | o4 | |
| HS0028+1747 | | 00 28 42.1 | +17 47 58 | 0.0978 | 18.1 | | E | o7 | |
| HS0029+1748 | NPM1G +17.0024 | 00 29 26.5 | +17 48 11 | 0.0073 | 17.5 | 18.03 | E | o7 | |
| HS0029+1443 | | 00 29 42.3 | +14 43 38 | 0.0175 | 17.2 | | E | o7 | |
| HS0032+0116 | UM 258 | 00 32 03.3 | +01 16 10 | 0.0147 | 19.0 | | E | o7 | |
| HS0033+0421 | UGC 00359 | 00 33 36.0 | +04 21 37 | 0.01637 | 16.5 | 15.5 | E | l19 | IrS |
| | UM 048 | | | | | | | | |
| | CGCG 409-047 | | | | | | | | |
| | CGCG 0033.6+0422 | | | | | | | | |
| | MCG +01-02-035 | | | | | | | | |
| | IRAS 00335+0421 | | | | | | | | |
| | IRAS F00336+0421 | | | | | | | | |
| HS0035+0725 | | 00 35 45.9 | +07 25 29 | 0.8541 | 18.3 | | E | o7 | QSO |
| HS0036+0437 | | 00 36 01.6 | +04 37 19 | 0.0289 | 17.3 | | E | o4 | |
| HS0036+0352 | | 00 36 05.7 | +03 52 32 | 0.3884 | 18.5 | | E | o5 | * |
| HS0037+0111 | | 00 37 45.2 | +01 11 18 | 0.0136 | 17.0 | | E | o4 | |
| HS0038+0122 | PC 0038+0122 | 00 38 06.1 | +01 22 52 | 0.066 | 17.6 | 18.11 | B | l67 | |
| HS0040+0952 | | 00 40 51.5 | +09 52 29 | 0.0156 | 17.3 | | E | o4 | |
| HS0041+2333 | | 00 41 41.3 | +23 33 36 | 0.0218 | 17.7 | | E | o4 | |
| HS0043+0531 | | 00 43 16.2 | +05 31 37 | 0.0408 | 17.6 | | E | o7 | |
| HS0044+0453 | UM 065 | 00 44 52.8 | +04 53 27 | 0.021 | 17.2 | | E | l77 | |
| HS0045+0020 | UM 279 | 00 45 55.6 | +00 20 50 | 0.0381 | 18.9 | | E | o7 | |
| HS0049-0006 | UM 282 | 00 49 13.4 | -00 06 33 | 0.037 | 19.1 | 18 | E | l77 | |
| | UCM 0049-0006 | | | | | | | | |
| HS0049+0017 | UM 283 | 00 49 15.5 | +00 17 38 | 0.01500 | 17.0 | 16.76 | E | l58 | |
| | UCM 0049+0017 | | | | | | | | |
| HS0051+0927 | | 00 51 22.4 | +09 27 27 | 0.0202 | 18.9 | | E | o7 | |
| HS0051+0555 | UM 077 | 00 51 44.4 | +05 55 35 | 0.017 | 18.6 | | B | l77 | |
| HS0052+2119 | | 00 52 33.0 | +21 19 52 | 0.0434 | 17.7 | | E | o4 | |
| HS0055+0104 | UM 293 | 00 55 19.8 | +01 04 04 | 0.0567 | 16.8 | 16.4 | E | l13 | Sy1.5 |
| HS0056+0044 | UM 295 | 00 56 21.3 | +00 44 09 | 0.0176 | 17.2 | | E | o4 | |
| | UCM 0056+0044 | | | | | | | | |
| HS0056+0043 | UM 296 | 00 56 30.0 | +00 43 55 | 0.01800 | 16.5 | 16.58 | E | l58 | IrS |
| | IRAS F00564+0043 | | | | | | | | |
| | UCM 0056+0043 | | | | | | | | |

Table 1. Continued

| (1) object name | (2) other names | (3) R.A. | (4) Decl. | (5) redshift | (6) $B_H$ | (7) $B_L$ | (8) S | (9) run or reference | (10) remarks |
|---|---|---|---|---|---|---|---|---|---|
| HS0058+1847 | | 00 58 52.6 | +18 47 25 | 0.0376 | 17.4 | | E | o7 | |
| HS0058+0638 | UM 082 | 00 58 44.7 | +06 38 42 | 0.051 | 17.2 | 17.88 | E | l77 | Sy 2 |
| | NPM1G +06.0048 | | | | | | | | |
| HS0101+0310 | | 01 01 24.9 | +03 10 56 | 0.0654 | 17.3 | | E | o4 | |
| HS0103+2441 | | 01 03 40.5 | +24 41 12 | 0.0541 | 17.1 | | B | o5 | |
| HS0103+1242 | IRAS F01037+1242 | 01 03 46.4 | +12 42 29 | 0.0447 | 16.5 | | E | o4 | IrS |
| HS0104+0622 | UM 085 | 01 04 09.3 | +06 22 00 | 0.041 | 17.2 | 17.29 | E | l77 | Sy 2 |
| | NPM1G +06.0054 | | | | | | | | |
| HS0106+1304 | | 01 06 14.5 | +13 04 14 | 0.0597 | 16.8 | | E | o4 | |
| HS0107+1946 | | 01 07 29.7 | +19 46 38 | 0.0423 | 18.0 | | E | o4 | |
| HS0107+2458 | | 01 07 58.5 | +24 58 26 | 0.0394 | 17.3 | | B | o5 | |
| HS0108+0150 | UM 306 | 01 08 00.4 | +01 50 52 | 0.0170 | 15.9 | | E | o4 | |
| HS0108+0103 | UM 307 | 01 08 56.6 | +01 03 19 | 0.0228 | 17.6 | 14.31 | E | l77 | HII region in an Sdm |
| | UGC 00749 | | | | | | | | IrS |
| | CGCG 385-022 | | | | | | | | |
| | CGCG 0109.0+0104 | | | | | | | | |
| | MCG +00-04-030 | | | | | | | | |
| | IRAS 01089+0103 | | | | | | | | |
| | IRAS F01089+0103 | | | | | | | | |
| HS0110+2149 | NPM1G +21.0055 | 01 10 31.6 | +21 49 54 | 0.0563 | 18.3 | 16.69 | E | o4 | |
| HS0111+1300 | UGC 00774 | 01 11 12.4 | +13 00 25 | 0.0491 | 16.8 | 14.76 | E | l13 | Sy 1 |
| | MRK 0975 | | | | | | | | IrS |
| | CGCG 0111.2+1300 | | | | | | | | Radio S |
| | IRAS F01112+1300 | | | | | | | | |
| | OC +118 | | | | | | | | |
| HS0111+2115 | NPM1G +21.0056 | 01 11 55.6 | +21 15 25 | 0.0318 | 16.3 | 16.42 | E | o7 | |
| HS0113+1750 | | 01 13 59.5 | +17 50 03 | 0.0626 | 18.5 | | E | o7 | |
| HS0115+1156 | MRK 0979 | 01 15 22.6 | +11 56 38 | 0.0190 | 15.6 | 15.5 | E | l44 | |
| HS0116+2244 | NPM1G +22.0058 | 01 16 53.0 | +22 44 33 | 0.0439 | 17.2 | 16.80 | B | o7 | abs |
| HS0117+1017 | | 01 17 13.6 | +10 17 33 | 0.0342 | 16.5 | | E | o6 | |
| HS0117+1135 | | 01 17 29.8 | +11 35 17 | 0.0615 | 16.2 | | B | o7 | |
| HS0118+1236 | | 01 18 36.4 | +12 36 49 | 0.0198 | 16.7 | | E | o6 | |
| HS0119+0331 | UM 099 | 01 19 32.6 | +03 31 41 | 0.0237 | 16.5 | | E | o4 | |
| HS0119+0044 | IRAS 01197+0044 | 01 19 43.9 | +00 44 42 | 0.0555 | 16.9 | | E | l72 | IrS |
| HS0122+0743 | UGC 00993 | 01 22 57.1 | +07 43 47 | 0.00975 | 17.6 | 15.00 | E | l19 | |
| HS0123+1624 | | 01 23 36.3 | +16 24 55 | 0.0290 | 17.2 | | E | o7 | |
| HS0124+1126 | NPM1G +11.0056 | 01 24 46.4 | +11 26 42 | 0.0327 | 16.7 | 17.36 | E | o6 | |
| HS0131+1937 | IRAS F01318+1936 | 01 31 52.3 | +19 37 04 | 0.0359 | 17.6 | | E | o7 | IrS |
| HS0133+1341 | | 01 33 43.0 | +13 41 43 | 0.0239 | 17.7 | | E | o7 | |
| HS0137+1539 | UGC 01176 | 01 37 30.4 | +15 39 34 | 0.0021 | 18.8 | 14.4 | E | o7 | HII region in an Im |
| | DDO 013 | | | | | | | | |
| | [RC2] A0137+15 | | | | | | | | |
| | LGG 029: [G93] 006 | | | | | | | | |
| HS0138+0458 | UM 126 | 01 38 48.8 | +04 58 04 | 0.032 | 17.4 | 18.0 | E | l80 | |
| HS0141+0719 | | 01 41 23.4 | +07 19 46 | 0.023 | 16.3 | | E | l80 | |
| HS0142+1651 | MRK 0361 | 01 42 03.7 | +16 51 31 | 0.0275 | 15.8 | 15.6 | E | l50 | |
| | CGPG 0142.0+1650 | | | | | | | | |
| | [RC2] A0142+16 | | | | | | | | |
| | NPM1G +16.0052 | | | | | | | | |
| HS0143+2400 | | 01 43 07.5 | +24 00 55 | 0.0346 | 18.4 | | E | o7 | |

Table 1. Continued

| (1) object name | (2) other names | (3) R.A. | (4) Decl. | (5) redshift | (6) $B_H$ | (7) $B_L$ | (8) S | (9) run or reference | (10) remarks |
|---|---|---|---|---|---|---|---|---|---|
| HS0143+0549 | UM 138 | 01 43 51.1 | +05 49 57 | 0.018 | 16.2 | 17.7 | E | l80 | |
| HS0148+1700 | | 01 48 32.0 | +17 00 13 | 0.0647 | 16.8 | | B | o7 | |
| HS0148+2123 | | 01 48 18.4 | +21 23 51 | 0.0165 | 17.2 | | E | o7 | |
| HS0153+2205 | | 01 53 17.9 | +22 05 05 | 0.0664 | 18.0 | | E | o7 | |
| HS0731+6348 | | 07 31 42.3 | +63 48 27 | 0.3447 | 18.6 | | B | o2 | Sy1 |
| HS0732+6503 | | 07 32 09.2 | +65 03 38 | 0.0217 | 17.2 | | B | o1 | |
| HS0737+6442 | | 07 37 35.2 | +64 42 18 | 0.036 | 17.7 | | E | l80 | |
| HS0746+6139 | KUG 0746+616 | 07 46 54.6 | +61 39 40 | 0.023 | 18.7 | | E | l80 | |
| HS0747+6456 | | 07 47 05.1 | +64 56 42 | 0.0247 | 16.5 | | E | o2 | |
| HS0749+5649 | | 07 49 37.7 | +56 49 48 | 0.0190 | 17.4 | | E | o7 | |
| HS0750+6019 | | 07 50 54.7 | +60 19 27 | 0.0356 | 17.8 | | E | o2 | |
| HS0752+5603 | | 07 52 45.1 | +56 03 07 | 0.0275 | 16.5 | | B | o7 | |
| HS0752+6147 | | 07 52 48.3 | +61 47 43 | 0.0287 | 16.0 | | E | o1 | |
| HS0757+6441 | | 07 57 19.2 | +64 41 52 | 0.0733 | 17.7 | | B | o5 | |
| HS0805+5742 | IRAS F08054+5742 | 08 05 25.6 | +57 42 30 | 0.0271 | 17.4 | | B | o7 | IrS |
| HS0808+5842 | SBS 0808+587 | 08 08 11.0 | +58 42 48 | 0.0272 | 17.7 | 15.5 | E | l43 | Sy 2 |
| | VII Zw 217 | | | | | | | | IrS |
| | CGCG 287-051 | | | | | | | | |
| | CGCG 0808.1+5843 | | | | | | | | |
| | CGPG 0808.1+5842 | | | | | | | | |
| | IRAS F08082+5842 | | | | | | | | |
| HS0814+6439 | | 08 14 47.8 | +64 39 03 | 0.0390 | 17.1 | | E | o2 | Sy1 |
| HS0831+6215 | | 08 31 18.4 | +62 15 43 | 0.0187 | 17.5 | | E | o7 | |
| HS0838+6253 | | 08 38 45.2 | +62 53 07 | 0.0044 | 16.0 | | E | o1 | |
| HS0847+6112 | MRK 0099 | 08 47 25.4 | +61 12 30 | 0.0125 | 16.2 | 16.6 | E | l44 | |
| | MCG +10-13-025 | | | | | | | | |
| | [RC2] A0847+61 | | | | | | | | |
| HS0912+5959 | MRK 0019 | 09 12 54.2 | +59 59 00 | 0.0141 | 16.7 | 16.0 | E | l45 | G pair, IrS |
| | CGCG 288-028 | | | | | | | | |
| | CGCG 0912.9+5959 | | | | | | | | |
| | MCG +10-13-071 | | | | | | | | |
| | IRAS 09129+5958 | | | | | | | | |
| | IRAS F09129+5958 | | | | | | | | |
| | [RC2] A0912+59 | | | | | | | | |
| HS0915+5540 | | 09 15 35.8 | +55 40 35 | 0.0494 | 17.2 | | E | o7 | |
| HS0930+5527 | MRK 0116 | 09 30 30.3 | +55 27 46 | 0.0031 | 16.2 | 15.6 | E | l45 | |
| | I Zw 018 | | | | | | | | |
| | CGPG 0930.5+5527 | | | | | | | | |
| | [RC2] A0930+55B | | | | | | | | |
| HS1222+3741 | CG 1022 | 12 22 08.2 | +37 41 13 | 0.0409 | 18.1 | | E | o7 | |
| HS1223+3938 | NPM1G +39.0289 | 12 23 29.8 | +39 38 30 | 0.0360 | 16.8 | 16.92 | E | o7 | |
| HS1232+3846 | | 12 32 15.5 | +38 46 56 | 0.0528 | 17.2 | | E | o7 | |
| HS1232+3947 | | 12 32 54.6 | +39 47 37 | 0.0210 | 17.2 | | E | o7 | |
| HS1232+3612 | KUG 1232+362 | 12 32 59.5 | +36 12 52 | 0.0425 | 16.2 | | E | o7 | G pair |
| | CG 1033 | | | | | | | | |
| HS1236+3821 | UGC 07816 | 12 36 31.8 | +38 21 51 | 0.0073 | 15.4 | | E | o7 | |
| | CGCG 188-012 | | | | | | | | |
| | CGCG 1236.5+3822 | | | | | | | | |
| | NPM1G +38.0259 | | | | | | | | |
| HS1244+3648 | | 12 44 37.0 | +36 48 05 | 0.0472 | 16.4 | | E | o7 | |

**Table 4: Continued**

| (1) object name | (2) other names | (3) R.A. | (4) Decl. | (5) redshift | (6) $B_H$ | (7) $B_L$ | (8) S | (9) run or reference | (10) remarks |
|---|---|---|---|---|---|---|---|---|---|
| HS1254+3323 | | 12 54 46.3 | +33 23 27 | 0.0032 | 17.7 | | B | o1 | abs |
| HS1255+3506 | | 12 55 22.2 | +35 06 32 | 0.0155 | 17.1 | | E | o2 | |
| HS1256+3505 | CG 1058 | 12 56 02.9 | +35 05 10 | 0.0342 | 16.5 | | E | o2 | |
| HS1256+3512 | | 12 56 51.2 | +35 12 19 | 0.0035 | 16.8 | | E | o2 | |
| HS1301+3312 | | 13 01 05.5 | +33 12 27 | 0.0371 | 17.8 | | E | o3 | |
| HS1301+3325 | | 13 01 20.0 | +33 25 37 | 0.0246 | 17.5 | | B | o2 | |
| HS1301+3209 | | 13 01 59.4 | +32 09 02 | 0.0238 | 17.8 | | E | o7 | |
| HS1302+3046 | | 13 02 35.4 | +30 46 30 | 0.0355 | 16.0 | | E | o3 | * |
| HS1304+3529 | CG 1085 | 13 04 03.7 | +35 29 43 | 0.0165 | 16.6 | | E | o3 | |
| HS1306+3525 | CG 1090 | 13 06 10.0 | +35 25 40 | 0.0165 | 16.6 | | E | o3 | |
| HS1306+3320 | | 13 06 12.8 | +33 20 39 | 0.0270 | 16.6 | | E | o3 | |
| HS1306+3527 | | 13 06 29.5 | +35 27 25 | 0.0371 | 17.0 | | E | o7 | |
| HS1308+3044 | CG 0984 | 13 08 42.5 | +30 44 55 | 0.0209 | 16.3 | | B | o7 | |
| HS1309+3409 | CG 1099 | 13 09 00.0 | +34 09 11 | 0.0785 | 17.0 | | B | o7 | |
| HS1309+3431 | KUG 1309+345B | 13 09 45.4 | +34 31 15 | 0.0168 | 16.2 | | E | o3 | spiral |
| HS1311+3628 | UGC 08303<br>DDO 166<br>CGCG 1311.0 +3628<br>MCG +06-29-061<br>IRAS F13110+3628<br>[RC2] A1310+36<br>[RC1] A1311<br>LGG 334:[G93] 004 | 13 11 00.0 | +36 28 02 | 0.0031 | 18.0 | 13.48 | E | o7 | IrS, HII region |
| HS1312+3847 | | 13 12 14.1 | +38 47 53 | 0.0515 | 17.3 | | E | o7 | |
| HS1312+3508 | | 13 12 28.3 | +35 08 47 | 0.0035 | 18.1 | | E | o3 | |
| HS1318+3406 | CG 1131 | 13 18 36.1 | +34 06 45 | 0.0352 | 16.6 | | E | o3 | |
| HS1319+3848 | NGC 5107<br>MRK 1346<br>KUG 1319+387<br>CGCG 217-033<br>CGCG 1319.1+3848<br>MCG +07-28-001<br>IRAS 13191+3847<br>IRAS F13191+3848<br>LGC 334:[G93] 011 | 13 19 08.3 | +38 48 05 | 0.00316 | 17.4 | 13.81 | E | l68 | IrS |
| HS1323+3211 | KUG 1323+321<br>[SMB88] 0540 | 13 23 16.8 | +32 11 23 | 0.0384 | 16.3 | | E | o3 | *, SBa(s) |
| HS1323+3319 | WAS 69<br>CG 1146 | 13 23 31.4 | +33 19 28 | 0.0153 | 16.7 | | E | o3 | |
| HS1323+3316 | MRK 0453<br>[SMB88] 0555<br>IRAS F13236+3316<br>KUG 1323+332<br>CG 1147 | 13 23 40.8 | 33 16 16 | 0.0465 | 16.4 | 16.5 | E | l44 | IrS |

| (1) object name | (2) other names | (3) R.A. | (4) Decl. | (5) redshift | (6) $B_H$ | (7) $B_L$ | (8) S | (9) run or reference | (10) remarks |
|---|---|---|---|---|---|---|---|---|---|
| HS1328+3132 | UGC 08602<br>CGCG 1328.3+3132<br>MCG +05-32-035<br>MRK 0455<br>VV 326a<br>[RC2] A1328+31<br>CGCG 161-074<br>[SMB88] 0767<br>CGCG 161-074E<br>KUG 1328+315 | 13 28 21.5 | +31 32 35 | 0.0342 | 15.3 | 14.6 | E | l14 | |
| HS1328+3424 | KUG 1328+344 | 13 28 44.5 | +34 24 35 | 0.0227 | 16.8 | | B | o7 | G pair |
| HS1329+3703 | | 13 29 52.7 | +37 03 39 | 0.0557 | 16.9 | | E | o7 | |
| HS1330+3651 | | 13 30 54.1 | +36 51 54 | 0.0167 | 16.7 | | E | o7 | |
| HS1331+3906 | KUG 1331+391 | 13 31 17.7 | +39 06 32 | 0.0643 | 16.4 | | B | o7 | |
| HS1332+3417 | MRK 0459<br>[RC2] A1332+34C<br>CG 1165 | 13 32 54.0 | +34 17 23 | 0.0241 | 15.7 | 17.0 | E | l44 | G pair |
| HS1333+3149 | | 13 33 06.8 | +31 49 36 | 0.0248 | 15.9 | | E | o3 | |
| HS1333+3058 | | 13 33 17.3 | +30 58 25 | 0.0402 | 17.3 | | E | o3 | |
| HS1334+3957 | | 13 34 11.5 | +39 57 32 | 0.0083 | 16.9 | | E | o7 | |
| HS1336+3114 | CGCG 1336.2+3115<br>CGCG 161-097<br>[SMB88] 1072<br>KUG 1336+312 | 13 36 12.5 | +31 14 26 | 0.0158 | 15.7 | | B | o7 | |
| HS1336+3650 | | 13 36 43.8 | +36 50 56 | 0.0202 | 17.1 | | E | o7 | |
| HS1338+3037 | CGCG 1338.9+3038<br>MRK 0268<br>CGCG1 161-119<br>IRAS F13389+3037<br>IRAS 13388+3037<br>KUG 1338+306B | 13 38 54.0 | +30 37 51 | 0.0410 | 16.2 | | E | l13 | G pair, IrS |
| HS1339+3046 | MRK 0067<br>UGCA 372<br>[RC2] A1339+30<br>IRAS F13396+3046<br>KUG 1339+307 | 13 39 39.5 | +30 46 17 | 0.0029 | 15.7 | 16.5 | E | l51 | IrS |
| HS1340+3307 | CG 1176 | 13 40 08.6 | +33 07 19 | 0.0158 | 16.8 | | E | o3 | |
| HS1340+3207 | | 13 40 28.3 | +32 07 59 | 0.0365 | 16.8 | | E | o3 | |
| HS1341+3409 | | 13 41 50.9 | +34 09 50 | 0.0171 | 17.3 | | E | o3 | |
| HS1344+3511 | IRAS 13448+3511<br>IRAS F13449+3511<br>CG 1189<br>[dKM92] | 13 44 54.0 | +35 11 35 | 0.0539 | 16.8 | 16 | E | l12 | IrS |
| HS1349+4027 | MRK 0462<br>KUG 1349+404<br>CGCG 218-064<br>CGCG 219-009<br>CGCG 1349.3+4027<br>MCG +07-29-002<br>[RC2] A1349+40<br>NPM1G +40.0336<br>LGG 361:[G93] 016 | 13 49 17.8 | 40 27 35 | 0.0079 | 17.0 | 15.38 | E | l44 | |
| HS1349+3942 | NPM1G +39.0332 | 13 49 22.9 | +39 42 03 | 0.0054 | 16.7 | 15.86 | E | o7 | |

Table 1. Continued

| (1) object name | (2) other names | (3) R.A. | (4) Decl. | (5) redshift | (6) $B_H$ | (7) $B_L$ | (8) S | (9) run or reference | (10) remarks |
|---|---|---|---|---|---|---|---|---|---|
| HS1353+3849 | MRK 0464 | 13 53 45.4 | +38 49 07 | 0.0510 | 16.9 | 16.5 | E | l13 | Sy 1.5, XrayS |
| | KUG 1353+388 | | | | | | | | |
| | H 1350+390 | | | | | | | | |
| | XRS 13505+390 | | | | | | | | |
| HS1354+3634 | CG 1200 | 13 54 27.5 | +36 34 28 | 0.0167 | 17.2 | | B | o7 | |
| HS1354+3635 | CG 1201 | 13 54 29.8 | +36 35 39 | 0.0171 | 16.1 | 15 | B | o7 | |
| HS1400+3927 | CG 0330 | 14 00 29.8 | +39 27 37 | 0.0045 | 17.2 | 17 | E | l78 | |
| | [SP82] 02 | | | | | | | | |
| HS1402+3657 | MRK 1369 | 14 02 06.6 | +36 57 53 | 0.0120 | 15.5 | 17.0 | E | l44 | IrS |
| | IRAS F14021+3657 | | | | | | | | |
| | CG 0340 | | | | | | | | |
| | NPM1G +36.0325 | | | | | | | | |
| HS1402+3650 | | 14 02 39.2 | +36 50 45 | 0.0347 | 16.3 | | E | o7 | |
| HS1408+4429 | CG 0368 | 14 08 23.6 | +44 29 01 | 0.0338 | 16.9 | 16 | B | l88 | |
| HS1408+4201 | | 14 08 39.7 | +42 01 06 | 0.0939 | 17.9 | | B | o7 | abs |
| HS1410+3627 | | 14 10 02.1 | +36 27 15 | 0.0338 | 16.9 | | E | o7 | |
| HS1410+3446 | MRK 0467 | 14 10 21.5 | +34 46 57 | 0.0316 | 15.8 | 16.5 | E | l44 | IrS |
| | [RC2] A1410+34 | | | | | | | | |
| | IRAS F14103+3447 | | | | | | | | |
| | KUG 1410+347 | | | | | | | | |
| | CG 0374 | | | | | | | | |
| HS1413+4402 | IRAS F14131+4402 | 14 13 08.2 | +44 02 07 | 0.0698 | 17.4 | | B | o7 | IrS |
| HS1413+3956 | KUG 1413+399 | 14 13 10.4 | +39 56 08 | 0.0426 | 16.1 | 16.5 | E | l6 | IrS |
| | IRAS F14131+3955 | | | | | | | | |
| | NPM1G +39.0344 | | | | | | | | |
| HS1415+4203 | | 14 15 57.1 | +42 03 05 | 0.0683 | 18.1 | | B | o7 | |
| HS1416+3554 | KUG 1416+359 | 14 16 03.3 | +35 54 27 | 0.0103 | 17.0 | 16 | E | o7 | spiral |
| HS1419+3639 | UGC 09198 | 14 19 16.2 | +36 39 19 | 0.0114 | 16.2 | 16.5 | E | l59 | Sb |
| HS1420+3437 | | 14 20 59.2 | +34 37 05 | 0.0246 | 18.2 | | E | o3 | |
| HS1421+4018 | | 14 21 37.9 | +40 18 43 | 0.0982 | 17.6 | | B | o7 | emi+abs |
| HS1422+3325 | | 14 22 18.9 | +33 25 58 | 0.0341 | 18.5 | | E | o3 | |
| HS1422+3339 | CG 0419 | 14 22 53.7 | +33 39 22 | 0.0114 | 17.1 | | E | o3 | |
| HS1425+3835 | CG 0435 | 14 25 14.6 | +38 35 33 | 0.0223 | 16.8 | 16 | E | o7 | |
| HS1429+3451 | CG 0457 | 14 29 14.3 | +34 51 17 | 0.0144 | 17.5 | 17.3 | E | o3 | |
| HS1429+3154 | CG 1236 | 14 29 35.4 | +31 54 39 | 0.0117 | 16.7 | 16.9 | E | o7 | |
| | NPM1G +31.0320 | | | | | | | | |
| HS1429+4511 | | 14 29 45.9 | +45 11 41 | 0.0321 | 18.0 | | E | o7 | |
| HS1435+4523 | | 14 35 06.7 | +45 23 03 | 0.1267 | 18.3 | | B | o7 | |
| HS1437+3701 | MRK 0475 | 14 37 03.6 | +37 01 12 | 0.0018 | 16.4 | 14.47 | E | l44 | |
| | CG 0493 | | | | | | | | |
| HS1438+3147 | CG 1250 | 14 38 33.7 | +31 47 39 | 0.0443 | 18.4 | 18 | E | o7 | |
| HS1440+3805 | NPM1G +38.0321 | 14 40 08.8 | +38 05 06 | 0.0322 | 16.7 | 17.00 | B | o7 | |
| HS1440+4302 | CG 0903 | 14 40 22.3 | +43 02 32 | 0.0085 | 17.6 | 17.00 | E | o7 | |
| | NPM1G +43.0283 | | | | | | | | |
| HS1442+4250 | | 14 42 17.9 | +42 50 13 | 0.0025 | 17.7 | | E | o7 | |
| HS1442+4332 | CG 0523 | 14 42 26.5 | +43 32 19 | 0.0811 | 17.5 | 16 | B | o7 | ** |
| HS1444+3114 | IRAS F14440+3114 | 14 44 01.6 | +31 14 23 | 0.0297 | 16.1 | 17 | E | o7 | IrS |
| | CG 1260 | | | | | | | | |
| HS1450+3844 | CG 0565 | 14 50 21.9 | +38 44 34 | 0.0140 | 17.8 | 16 | E | l78 | |
| | [SP82] 25 | | | | | | | | |
| HS1502+4152 | | 15 02 31.3 | +41 52 35 | 0.0164 | | | B | o7 | |

| (1) object name | (2) other names | (3) R.A. | (4) Decl. | (5) redshift | (6) $B_H$ | (7) $B_L$ | (8) S | (9) run or reference | (10) remarks |
|---|---|---|---|---|---|---|---|---|---|
| HS1504+3922 | CG 0624<br>[SP82] 31 | 15 04 15.5 | 39 22 15 | 0.0302 | 18.8 | 17 | E | l78 | |
| HS1505+3944 | CG 0632 | 15 05 53.6 | +39 44 19 | 0.0366 | 17.6 | 16 | E | o7 | |
| HS1507+3743 | CG 0644 | 15 07 38.1 | +37 43 06 | 0.0322 | 18.3 | 18 | E | o7 | |
| HS1522+4214 | CG 0702<br>NPM1G +42.0413 | 15 22 23.3 | +42 14 40 | 0.0190 | 17.1 | | B | o7 | |
| HS1524+4205 | | 15 24 08.0 | +42 05 01 | 0.0225 | 18.4 | | E | o7 | |
| HS1526+4045 | | 15 26 56.7 | +40 45 17 | 0.0288 | 17.7 | | B | o7 | |
| HS1543+4525 | | 15 43 23.3 | +45 25 45 | 0.0389 | 17.4 | | E | o3 | |
| HS1544+4736 | | 15 44 28.0 | +47 36 20 | 0.0195 | 18.0 | | E | o3 | |
| HS1546+4755 | | 15 46 56.3 | +47 55 34 | 0.0377 | 18.9 | | E | o3 | |
| HS1548+4745 | PC 1548+4745 | 15 48 04.2 | +47 45 09 | 0.070 | 18.9 | | B | l67 | |
| HS1549+4630 | PC 1549+4630 | 15 49 35.5 | +46 30 37 | 0.098 | 18.9 | | B | l67 | |
| HS1609+4827 | | 16 09 44.4 | +48 27 44 | 0.0096 | 16.4 | | E | o3 | |
| HS1610+4539 | | 16 10 40.9 | +45 39 37 | 0.0196 | 17.7 | | E | o3 | |
| HS1614+4709 | | 16 14 54.2 | +47 09 22 | 0.0026 | 16.9 | | E | o3 | |
| HS1626+5153 | MRK 1498<br>IRAS F16268+5152<br>IRAS 16268+5152<br>[dKM92] 408 | 16 26 48.4 | +51 53 05 | 0.0547 | 16.8 | 17.0 | E | l12 | Sy 1, IrS |
| HS1627+5239 | | 16 27 33.1 | +52 39 36 | 0.0288 | 18.3 | | E | o3 | |
| HS1633+4703 | | 16 33 12.7 | +47 03 44 | 0.0086 | 16.7 | | E | o3 | |
| HS1634+5218 | CGCG 1634.0+5220<br>MRK 1499<br>UGC A 412<br>CGCG 276-029<br>[RC2] A1634+52<br>IRAS F16340+5219<br>CGPG 1634.0+5220<br>I Zw 159 | 16 34 07.7 | +52 18 56 | 0.0087 | 15.0 | 16.0 | E | l44 | G pair, IrS |
| HS1640+5136 | CGCG 1640.8+5138<br>MRK 1500<br>CGCG 276-037<br>IRAS F16408+5136<br>IRAS 16407+5136 | 16 40 48.2 | +51 36 30 | 0.0308 | 15.8 | 15.6 | E | o3 | IrS |
| HS1641+5053 | IRAS F16415+5053 | 16 41 31.6 | +50 53 22 | 0.0292 | 16.4 | | E | o3 | IrS |
| HS1643+5313 | | 16 43 10.2 | +53 13 16 | 0.785 | 19.5 | | E | o3 | QSO |
| HS1645+5155 | | 16 45 19.8 | +51 55 42 | 0.0286 | 19.5 | | E | o3 | |
| HS1657+5033 | CGCG 1657.3+5034<br>CGCG 252-012<br>IRAS F16573+5034 | 16 57 22.2 | +50 33 53 | 0.0102 | 16.0 | | E | o3 | IrS |
| HS1657+5735 | MRK 0891<br>IRAS F16576+5735<br>IRAS 16576+5735<br>CGPG 1657.6+5736<br>Zw 670 | 16 57 36.2 | +57 35 48 | 0.0505 | 16.2 | | E | o3 | IrS |
| HS1711+5758 | | 17 11 38.2 | +57 58 34 | 2.996 | 19.0 | | E | o7 | QSO |
| HS1723+5631 | IRAS F17237+5631 | 17 23 43.5 | +56 31 14 | 0.0286 | 17.0 | | E | o3 | IrS |
| HS1728+5655 | | 17 28 14.4 | +56 55 36 | 0.0160 | 17.0 | | E | o3 | |
| HS1734+5704 | IRAS F17341+5704 | 17 34 10.1 | +57 04 58 | 0.0475 | 16.3 | | E | o3 | IrS |
| HS2353+2005 | | 23 53 39.4 | +20 05 13 | 0.0236 | 17.5 | | E | o4 | |

* - galaxies with only one line detection
** - galaxies with only one line detection but twice observed

**List of Abbreviations**

| abrrev. | | reference |
|---|---|---|
| A | Abell galaxy | 1 |
| CG | Case Galaxy | 53-57, 62-65, 71 |
| CGCG | Catalogue of Galaxies and of Cluster of Galaxies 1968 | 93 |
| CGPG | Catalogue of Selected Compact Galaxies and of Post Eruptive Galaxies 1971 | 92 |
| DDO | David Dunlap Observatory catalog, van den Bergh 1966 | 3 |
| [dKM92] | de Grijp, Keel, Miley et. al. 1992 | 12 |
| [G93] | Garcia '93 | 11 |
| H | Hard X-Ray Sources | 42 |
| IRAS | Infrared Astronomical Satellite Catalogs 1988. The Point Source Catalog | 15 |
| IRAS F | Infrared Astronomical Satellite Catalogs 1990. The Faint Source Catalog | 46 |
| KUG | Kiso Survey for UV Excess Galaxies | 73-76 |
| LGG | Lyon Group of Galaxies Catalog | 11 |
| MCG | Morphological Catalogue of Galaxies 1962 | 82-86 |
| MRK | Markarian Galaxy | 25-40 |
| NPM | Lick Northern Proper Motion Program | 16 |
| OA-OZ | Ohio Source Catalog, 1415 MHz | 17, 47, 18, 66, 7, 10, 8, 5, 9, 60 |
| PC | Palomar Transit Grism Survey | 67 |
| [RC2] | The Second Reference Catalogue 1976 | 79 |
| [SMB88] | Slezak, Mars, Bijaoi et. al. 1988 | 69 |
| SBS | Second Byurakan Spectral Sky Survey | 41, 70 |
| [SP82] | Sanduleak and Pesch 1982 | 61 |
| UCM | Universidad Complutense de Madrid | 58, 89, 90 |
| UGC | Uppsala General Catalogue of Galaxies 1973 | 48 |
| UM | University of Michigan | 20-24 |
| VV | Vorontsov-Velyaminov 1959, 1977, Atlas and Catalog of Interacting Galaxies | 81 |
| WAS | Wasilewski 1983 | 87 |
| XRS | Second Catalogue of X-ray Sources | 2 |
| Zw | Zwicky Compact Galaxy | 91 |

1. Abell, G.O., Corwin, H.G., and Olowin, R.P. 1989, ApJS 70, 1
2. Amnuel, P.R., Guseinov, O.H., and Rakhamimov, Sh. Yu. 1982, Astrophys. and Space Sci. 82, 3
3. van den Bergh, S. 1966, AJ 71, 922
4. Bothun, G.D., Halpern, J.P., Lonsdale, C.J., Impey, C., and Schmitz, M. 1989, ApJS 70, 271
5. Brundage, R.K., Dixon, R.S., Ehman, J.R., and Kraus, J.D. 1971, AJ 76, 777
6. Dey, A., and Strauss, A. 1990, AJ 99, 463
7. Dixon, R.S., and Kraus, J.D. 1968, A.J. 73, 381
8. Ehman, J.R., Dixon, R.S., and Kraus, J.D. 1970, AJ 75, 351
9. Ehman, J.R., Dixon, R.S., Ramakrishna, C.M., and Kraus, J.D. 1974, AJ 79, 144
10. Fitch, L.T., Dixon, R.S., and Kraus, J.D. 1969, AJ 74, 612
11. Garcia, A.M. 1993, A&AS 100, 46
12. de Grijp, M.H.K., Keel, W.C., Miley, G.K., Goudfrooij, P., and Lub, J. 1992, A&AS 96, 389
13. Hewitt, A., and Burbidge, G. 1991, ApJS 75, 297
14. Huchra, J.P., Geller, M.J., de Lapparent, V., and Corwin JR, H.G. 1990, ApJS 72, 433
15. Joint IRAS Science Working Group 1988, Infrared Astronomical Satellite Catalogs, 1988, The Point Source Catalog, Version2.0, NASA RP-1190
16. Klemola, A.R., Jones, B.F., and Hanson, R.B. 1987, AJ 94, 501
17. Kraus, J.D. 1964, Nature 202, 269
18. Kraus, J.D., Dixon, R.S., and Fisher, R.O. 1966, Ap.J. 144, 559
19. Lu, N.Y., Hoffman, G.L., Groff, T., Ross, T., and Lamphier, C. 1993, ApJS 88, 383
20. MacAlpine, G.M., Smith, S.B., and Lewis, D.W. 1977, ApJS 34, 95 (List I)
21. MacAlpine, G.M., Smith, S.B., and Lewis, D.W. 1977, ApJS, 35, 197 (List II))
22. MacAlpine, G.M., Lewis, D.W., Smith, S.B. 1977, ApJS 35, 203 (List III)
23. MacAlpine, G.M., Lewis, D.W. 1978, ApJS 36, 587 (List IV)
24. MacAlpine, G.M., and Williams, G.A. 1981, ApJS 45, 1113 (List V)
25. Markarian, B.E.,1967, Astrofizika 3, 55
26. Markarian, B.E., 1969, Astrofizika 5, 443
27. Markarian, B.E., and Lipovetskii, V.A. 1969, Astrofizika 5, 581
28. Markarian, B.E., and Lipovetskii, V.A. 1971, Astrofizika 7, 511
29. Markarian, B.E., and Lipovetskii, V.A. 1972, Astrofizika 58, 155
30. Markarian, B.E., and Lipovetskii, V.A. 1973, Astrofizika 9, 487
31. Markarian, B.E., and Lipovetskii, V.A. 1974, Astrofizika 10, 307
32. Markarian, B.E., and Lipovetskii, V.A. 1976a, Astrofizika 12, 389
33. Markarian, B.E., and Lipovetskii, V.A. 1976b, Astrofizika 12, 657
34. Markarian, B.E., Lipovetskii, V.A., and Stepanian, D.A. 1977a, Astrofizika 13, 225
35. Markarian, B.E. Lipovetskii, V.A., and Stepanian, D.A. 1977b, Astrofizika 13, 397
36. Markarian, B.E. Lipovetskii, V.A., and Stepanian, D.A. 1979a, Astrofizika 15, 201
37. Markarian, B.E. Lipovetskii, V.A., and Stepanian, D.A. 1979b, Astrofizika 15, 363
38. Markarian, B.E. Lipovetskii, V.A., and Stepanian, D.A. 1979c, Astrofizika 15, 54
39. Markarian, B.E. Lipovetskii, V.A., and Stepanian, D.A. 1981, Astrofizika 17,61
40. Markarian, B.E., Lipovetskii, V.A., and Stepanian, D.A. 1983, Astrofizika 19, 221
41. Markarian, B.E., and Stepanian, J.A. 1983, Astrophysics 19, 354; Astrofizika 19,639
42. Marshall, F.E., Boldt, E.A., Holt, S.S., Mushotzky, R.F., Pravdo, S.H., Rothschild, R.E., and Serlemitsos, P.J. 1979, ApJS 40, 657
43. Martel, A., and Osterbrock, D.E. 1994, AJ 107, 1283
44. Mazzarella, J.M., and Balzano, V.A. 1986, ApJS 62, 751
45. Mazzarella, J.M., and Boroson, T.A. 1993, ApJS 85, 27
46. Moshir, M., Kopan, G., Conrow, T., Mccallon, H., Hacking, P., Gregorich, D., Rohrbach, G., Melnyk, M., Rice, W. Fullmer, L., et. al. 1990, Infrared Astronomical Satellite Catalogs, 1990, The Faint Source Catalog, Version 2.0
47. Nash, R.T. 1965, AJ 70, 846
48. Nilson, P. 1973, Uppsala General Catalogue of Galaxies, Nova Acta Regiae Societatis Scientiarum Ser. V, Vol.1, Uppsala
49. Noguchi, T., Maehara, H., and Kondo, M. 1980, Ann. Tokio Astron. Obs., 2-nd Series 18, 55
50. Osterbrock, D.E., and Pogge, R.W. 1987, ApJ 323, 108
51. Osterbrock, D.E., and Shaw, R.A. 1988, ApJ 327, 89
52. Palumbo, G.C.G., Tanzella-Nitti, G., and Vettolani, G. 1983, Catalog of Radial Velocities of Galaxies, New-York: Gordon&Breach
53. Pesch, P., Sanduleak, N. (Paper I) 1983, ApJS 51, 171
54. Pesch, P., Sanduleak, N. (Paper III) 1986, ApJS 60, 543
55. Pesch, P., Sanduleak, N. (Paper V) 1988, ApJS 66, 297
56. Pesch, P., Sanduleak, N. (Paper VIII) 1989, ApJS 70, 163
57. Pesch, P., Sanduleak, N., and Stephenson, C.B. (Paper XII) 1991, ApJS 76,1043
58. Rego, M., Cordero-Gracia, M., Zamorano, J. and Gallego, J. 1993, AJ 105, 427
59. Richter, O.-G., and Huchtmeier, W.K. 1991, A&AS 87, 425
60. Rinsland, C.P., Dixon, R.S., Gerhart, M.R., and Kraus, J.D. 1974, AJ 79, 1129
61. Sanduleak, N. and Pesch, P. 1982, ApJ 258, L11
62. Sanduleak, N., Pesch, P. (Paper II) 1984, ApJS 55, 517
63. Sanduleak, N., and Pesch, P.(Paper IV) 1987, ApJS 63, 809
64. Sanduleak, N., Pesch, P. (Paper IX) 1989, ApJS 70, 173

**References to the Table 4. Continued**

65. Sanduleak, N., Pesch, P. (Paper XI) 1990, ApJS 72, 291
66. Scheer, D.J., and Kraus, J.D. 1967, A.J. 72, 536
67. Schneider, D.P., Schmidt, M., and Gunn, J.E., 1994, AJ 107, 1245
68. Schneider, S.E., Thuan, T.X., Mangum, J.G. and Miller, J. 1992, ApJS 81, 5
69. Slezak, E., Mars, G., Bijaoui, A., Balkowski, C., and Fontanelli, P. 1988, A&AS 74, 83
70. Stepanian, J.A., Lipovetski, V.A. and Erastova, L.K. 1990, Astrophysics 32, 252; Astrofizika 32, 441
71. Stephenson, C.B., Pesch, P., and MacConnell, D.J. (Paper XIII) 1992, ApJS 82, 471
72. Strauss, M.A., Huchra, J.P., Davis, M., Yahil, A., Fisher, K.B., and Tonry, J. 1992, ApJS 83, 29
73. Takase, B., and Miyauchi-Isobe, N. 1984, Annals of the Tokio Ast. Obs. 19, 595
74. Takase, B., and Miyauchi-Isobe, N. 1986, Annals of the Tokio Ast. Obs. 21, 181
75. Takase, B., and Miyauchi-Isobe, N. 1987, Annals of the Tokio Ast. Obs. 21, 251
76. Takase, B, and Miyauchi-Isobe, N. 1989 Pub. National Ast. Obs. of Japan 1, 11
77. Terlevich, R., Melnick, J., Masegosa, J., Moles, M., and Copetti, M.V.F. 1991, A&AS 91, 285
78. Tifft, W.C., Kirshner, R.P., Gregory, S.A., Moody, J.W. 1986, ApJ 310,75
79. de Vaucouleurs, G., de Vaucouleurs, A., and Corwin, H.G. 1976, Second Reference Catalogue of Bright Galaxies, University of Texas Press, Austin
80. Vogel, S., Engels, D., Hagen, H.-J., Groote, D., Wisotzki, L., Cordis, L., and Reimers, D. 1993, A&AS 98, 193
81. Vorontsov-Velyaminov, B.A. 1959, Atlas and Catalog of Interacting Galaxies, Sternberg. Inst., Moscow State University
82. Vorontsov-Velyaminov, B.A., and Krasnogorskaja, A.A. 1962, Morphological Catalog of Galaxies, Part I Moscow State University
83. Vorontsov-Velyaminov, B.A., and Arhipova, V.P. 1964, Morphological Catalog of Galaxies, Part II Moscow State University
84. Vorontsov-Velyaminov, B.A., and Arhipova, V.P. 1963, Morphological Catalog of Galaxies, Part III Moscow State University
85. Vorontsov-Velyaminov, B.A., and Arhipova, V.P. 1968, Morphological Catalog of Galaxies, Part IV Moscow State University
86. Vorontsov-Velyaminov, B.A., and Arhipova, V.P. 1974, Morphological Catalog of Galaxies, Part V Moscow State University
87. Wasilewski, A.J. 1983, ApJ 272, 68
88. Weistrop, D., and Downes, R.A. 1988, ApJ 331, 172
89. Zamorano, J., Rego, M., Gonzáles-Riestra, R. 1989, 79, 443
90. Zamorano, J., Rego, M., Gallego, J., Vitores, A.G. , Gonza'ales-Riestra, R., and Rodrígues-Caderot, G.
91. Zwicky, F. 1961-1968, Seven privately circulated lists.
92. Zwicky, F. 1971, Catalogue of Selected Compact Galaxies and of Post Eruptive Galaxies
93. Zwicky, F., Herzog, E., Wild, P., Karpowicz, M., and Kowal, C. 1961-1968, Catalog of Galaxies and of Clusters of Galaxies, Pasadena: California Institute of Technology

## 4. Results

Complete follow-up spectroscopy of our first-priority candidates was accomplished. With a success rate of detecting true emission-line objects of 76%, this provide the bulk of our sample of ELGs. As mentioned in § 2, we also observed 83 second priority candidates, in order to test their nature. From these, many were stars, but we also found 23 emission line galaxies, one galaxy with emission and absorption, four galaxies with absorption lines and a QSO. The efficiency of finding emission line objects in the second priority sample is then 28%. This result is biased, because we observed mostly candidates that showed extended images on the direct plates. From the 23 emission line galaxies we found, 21 had obvious extended images, and only two looked like point sources. If one takes random candidates from the second priority list, the efficiency is as low as 10%. From the 23 galaxies with emission, only two had intermediate ionization spectra that could be detected in the IIIa-J objective plates. The others showed either low or very low ionization spectra, with faint $H_\beta$ and [OIII], or spectra with only $H_\alpha$, or $H_\alpha$ and [SII]. These objects cannot be seen as emission-line candidates in the IIIa-J objective plate. We conclude then that there is no way to distinguish real emission-line galaxies from the rest of non-emission blue candidates. Taking into account the low efficiency of finding emission-line objects, such a blind search would cost too much telescope time. We believe that such objects can be very easily recognised in an IIIa-F survey ( like that of Zamorano et al. 1994). Therefore, we do not intend to complete the follow-up spectroscopy for the second priority objects, and we leave them to the discovery of the $H\alpha$ surveys.

We obtained a final sample of 203 objects, of which 196 are emission-line galaxies, four are galaxies with absorption, and three are QSOs. From our sample of 203 objects, 98 (48%) are newly discovered objects, 52 (26%) are objects found already in the literature but with no available redshift (sometimes mentioned only as IRAS sources or only as NPM (Lick Northern Proper Motion Program) galaxies) and 53 (26%) are objects with redshifts given already in the literature. We have observed all the objects with unknown redshift. From the objects with available redshifts, five were reobserved, in order to estimate our redshift errors (see Section 3). The mean surface density of the emission-line galaxies of our survey is $0.16\,\mathrm{deg}^{-2}$. This number is quite low in comparison with the mean surface density of 0.46, found by Salzer et al. (1989). But one must keep in mind that this is a sample of candidates selected in a special parameter region. We select our candidates in a certain "brightness"range and in a certain "colour" range. We are also losing the low-ionization galaxies into the second category objects, as discussed above. Thus our sample is dominated by high ionization galaxies, which should not come as a surprise, since the final selection was on the [OIII] line strengths.

The apparent magnitudes (derived from the Hamburg prism plates) of our sample range between $15.0 \leq B \leq 19.5$. We have selected all available magnitudes (56) from the literature for our sample (Table 4) and compare this heterogeneous data set with our own magnitude estimates (Fig. 5). The values agree sufficiently well below 16 mag, if we assume internal errors of about 0.5 mag in each sample. For the literature values, 0.5 mag is an estimated mean error, as these data came from various sources, with different measurement methods and calibrations. The estimated error for the prism plate magnitudes is known to be 0.5 mag. For galaxies with $B \leq 16$, our magnitudes seem to underestimate the total flux of the galaxies systematically. This can easily be explained by the fact that the objective prism spectra show up only with the bright HII region (or the core of the galaxy), while the fainter surrounding, without emission lines, is not detectable on these plates (see remarks in Table 4). A typical example is HS0108+0103, an Sdm galaxy, that we see only as an HII knot. The apparent magnitude of 17.6 describes only the flux that comes from the HII region, as compared with the value of $14.31^m$ of the whole galaxy. Another example is HS0137+1539, a very compact source near the edge of the low-surface brightness Im galaxy, DDO 013. With B=18.8, as compared with the B=14.4 apparent magnitude of the DDO, it was not obvious if the candidate was a separate object, or part of the DDO galaxy. A long slit spectra of the object confirmed that it belongs to the DDO galaxy. The same situation happened with HS1311+3628. The galaxy was reobserved in order to see if the emission knot we found is a separate galaxy or an HII region in the DDO 166. As the redshifts agree, the source we found is an HII region (apparent magnitude 18.0 instead of 13.48). Other galaxies where we subestimate the contribution of the underlying galaxies are: HS0122+0743 (B=17.6 instead of B=15.00), HS0808+5842 (B=17.7 instead of B=15.5), HS1332+3417 (B=15.7 instead of B=17.0), HS1349+4027 (B=17.0 instead of 15.38), HS1437+3701 (B=16.4 instead of B=14.47).

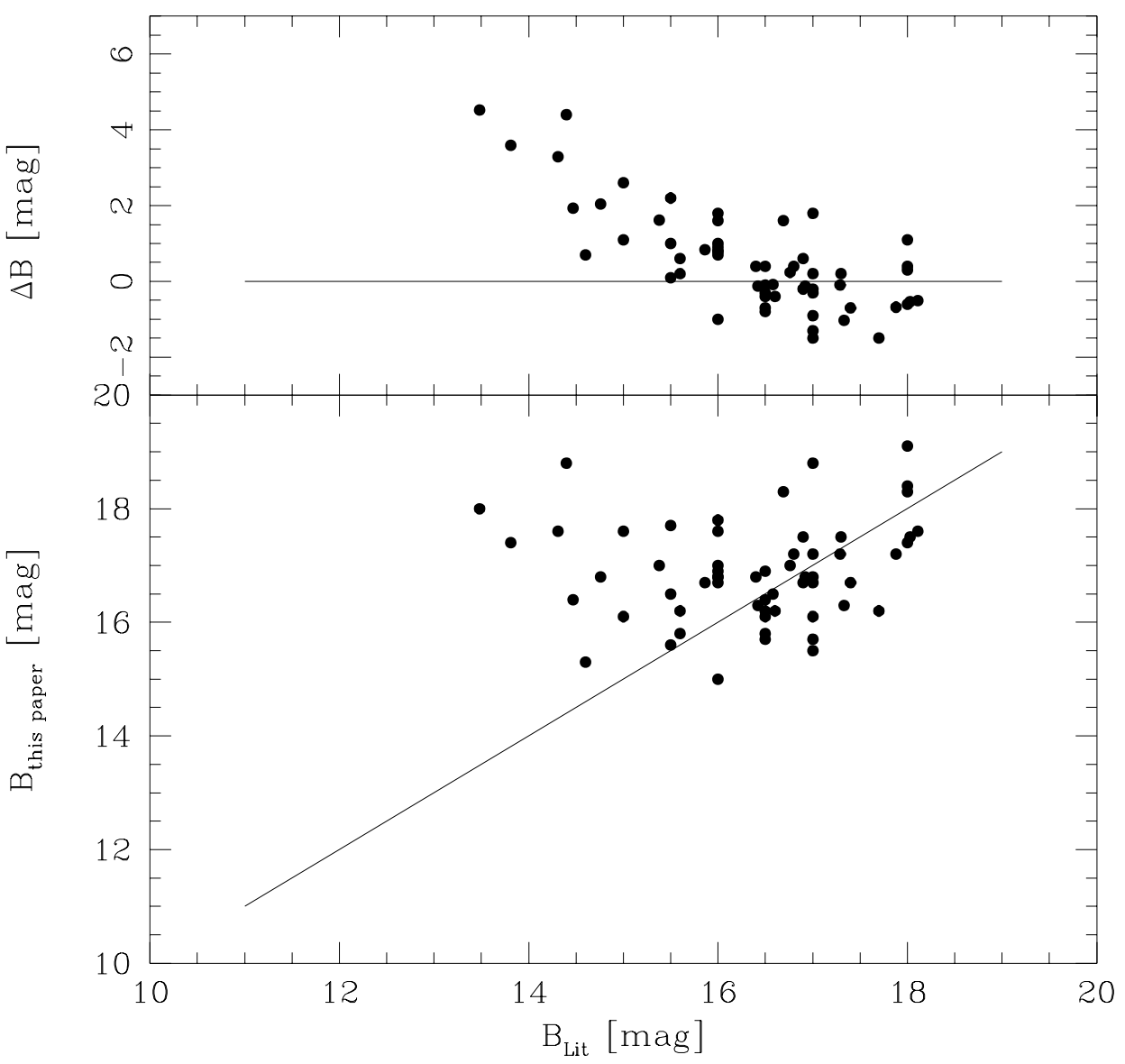

**Fig. 5.** The comparison between our Hamburg objective prism magnitudes $B_H$ and the magnitudes from literature, $B_L$. In the bottom panel we plotted $B_H$ versus $B_L$. The solid line represents $B_H = B_L$. In the top panel we plotted the difference $\Delta B = B_H - B_L$ versus $B_L$.

To have a further independent comparison of our magnitudes, we used the on-line facility of the APM catalogue as

**Table 5: The emission-lines detected in the spectra of the observed objects**

| object name | emission lines |
|---|---|
| HS0000+2252 | [OII], $H_\delta$, $H_\gamma$, $H_\beta$, [OIII], [OIII], HeI, $H_\alpha$+[NII], HeI, [SII] |
| HS0000+2422 | [OII], $H_\beta$, [OIII], [OIII], $H_\alpha$, [SII] |
| HS0013+0809 | [OIII], $H_\alpha$ |
| HS0013+1942 | Mgb, NaI, $H_\alpha$ |
| HS0013+2241 | [OII], $H_\beta$, [OIII], [OIII], [OI]?, $H_\alpha$+[NII], [SII] |
| HS0016+1449 | [OII], $H_\beta$, [OIII], [OIII], $H_\alpha$, [SII] |
| HS0021+1347 | [OII], $H_\beta$, [OIII], [OIII], $H_\alpha$, [SII], |
| HS0022+0014 | [OII], $H_\beta$, [OIII], [OIII], HeI, $H_\alpha$+[NII], [SII] |
| HS0024+1022 | [OII], $H_\beta$, [OIII], [OIII], $H_\alpha$, [SII], |
| HS0022+1945 | $H_\alpha$, [SII] |
| HS0024+2314 | [OII], $H_\beta$, [OIII], [OIII], $H_\alpha$, [SII] |
| HS0026+0332 | [OII], $H_\beta$, [OIII], [OIII], HeI?, $H_\alpha$, [SII] |
| HS0028+1747 | $H_\beta$, $H_\alpha$ |
| HS0029+1748 | [NeIII], $H_\zeta$, $H_\epsilon$, $H_\delta$, $H_\gamma$, [OIII], $H_\beta$, [OIII], [OIII], HeI, $H_\alpha$+[NII], HeI, [SII], [AIII], [OII] |
| HS0029+1443 | [OII], [NeIII], $H_\zeta$, $H_\epsilon$, $H_\delta$, $H_\gamma$, [OIII], $H_\beta$, [OIII], [OIII], HeI, [OI] $H_\alpha$+[NII], HeI, [SII], [AIII], [OII], [AIII] |
| HS0035+0725 | QSO: MgII |
| HS0036+0437 | [OII], $H_\beta$, [OIII], [OIII], $H_\alpha$, [SII] |
| HS0036+0352 | [OII] ? |
| HS0037+0111 | [OII], $H_\beta$, [OIII], [OIII], $H_\alpha$+[NII], [SII] |
| HS0040+0952 | [OII], $H_\gamma$, $H_\beta$, [OIII], [OIII], [OI]?, $H_\alpha$, [SII] |
| HS0041+2333 | [OII], [NeIII], $H_\zeta$, [NeIII]+[OII]+$H_\epsilon$?, $H_\delta$, $H_\gamma$, $H_\beta$, [OIII], [OIII], HeI, $H_\alpha$, HeI?, [SII] |
| HS0043+0531 | [OII], $H_\beta$, [OIII], [OIII], $H_\alpha$+[NII], [SII] |
| HS0045+0020 | $H_\beta$, [OIII], [OIII], $H_\alpha$, [SII] |
| HS0051+0927 | $H_\beta$, [OIII], [OIII], HeI, $H_\alpha$+[NII], [SII] |
| HS0052+2119 | [OII], $H_\beta$, [OIII], [OIII], HeI?, $H_\alpha$, [SII] |
| HS0056+0044 | [OII], $H_\beta$, [OIII], [OIII], $H_\alpha$, [SII] |
| HS0058+1847 | [OII], [NeIII], $H_\epsilon$, $H_\delta$, $H_\gamma$, $H_\beta$, [OIII], [OIII], HeI, [OI], $H_\alpha$+[NII], [SII], [AIII] |
| HS0101+0310 | [OII], [NeIII]?, $H_\gamma$?, $H_\beta$, [OIII], [OIII], HeI?, $H_\alpha$, [SII] |
| HS0103+2441 | [OII], $H_\beta$, [OIII], [OIII], $H_\alpha$, [SII] |
| HS0103+1242 | [OII], $H_\gamma$?,$H_\beta$, [OIII], [OIII], HeI?, $H_\alpha$+[NII], [SII] |
| HS0106+1304 | [OII], [NeIII], $H_\beta$, [OIII], [OIII], [OI], $H_\alpha$+[NII], [SII] |
| HS0107+1946 | $H_\beta$, [OIII], $H_\alpha$+[NII], [SII] |
| HS0107+2458 | [OII], $H_\beta$, [OIII], [OIII], $H_\alpha$, [SII] |
| HS0108+0150 | $H_\beta$, [OIII], [OIII], $H_\alpha$, [SII] |
| HS0110+2149 | $H_\beta$, [OIII], [OIII], $H_\alpha$+[NII], [SII] |
| HS0111+2115 | [OII], [NeIII], $H_\zeta$, $H_\epsilon$, $H_\delta$, $H_\gamma$, $H_\beta$, [OIII], [OIII], HeI, [OI], $H_\alpha$, [SII], [AIII] |
| HS0113+1750 | [OII], [NeIII], $H_\zeta$, $H_\epsilon$, $H_\delta$, $H_\gamma$, [OIII], $H_\beta$, [OIII], [OIII], HeI, [OI], $H_\alpha$, HeI, [SII], [AIII] |
| HS0116+2244 | abs: $H_\beta$, Mgb, NaD |
| HS0117+1017 | [OII], $H_\gamma$, [OIII], [OIII] |
| HS0117+1135 | [NII]+$H_\alpha$+[NII], [SII] |
| HS0118+1236 | [OII], $H_\gamma$, [OIII], [OIII] |
| HS0119+0331 | [OII], $H_\beta$, [OIII], [OIII], HeI?, [OI]?, $H_\alpha$+[NII], [SII] |
| HS0123+1624 | $H_\gamma$, $H_\beta$, [OIII], [OIII], HeI, [OI], $H_\alpha$, [SII] |
| HS0124+1126 | [OII], [OIII], [OIII] |
| HS0131+1937 | $H_\beta$, [OIII], [OIII], $H_\alpha$+[NII], [SII] |
| HS0133+1341 | $H_\beta$, [OIII], [OIII], $H_\alpha$+[NII], [SII] |
| HS0137+1539 | $H_\beta$, [OIII], [OIII], $H_\alpha$+[NII], [SII] |
| HS0143+2400 | [OII], [NeIII], $H_\zeta$, $H_\epsilon$, $H_\delta$, $H_\gamma$, [OIII], $H_\beta$, [OIII], [OIII], HeI, $H_\alpha$, HeI, [SII], [AIII] |
| HS0148+1700 | [OII], $H_\gamma$, $H_\beta$, [OIII], [OIII], [OI], $H_\alpha$+[NII], [SII] |

**Table 5: Continued**

| object name | emission lines |
|---|---|
| HS0148+2123 | $H_\delta$, $H_\gamma$, $H_\beta$, [OIII], [OIII], HeI, $H_\alpha$+[NII], [SII], [AIII] |
| HS0153+2205 | [OIII], [NII]+$H_\alpha$+[NII], [SII] |
| HS0731+6348 | Sy1: MgII, $H_\gamma$, $H_\beta$, [OIII], [OIII], $H_\alpha$ |
| HS0732+6503 | $H_\alpha$+[NII], [SII] |
| HS0747+6456 | [OII], [NeIII], $H_\beta$, [OIII], [OIII], [OI], $H_\alpha$, [SII], |
| HS0749+5649 | [NeIII], $H_\zeta$, $H_\epsilon$, $H_\delta$, $H_\gamma$, [OIII], $H_\beta$, [OIII], [OIII], HeI, [OI], $H_\alpha$+[NII], HeI, [SII], [AIII] |
| HS0750+6019 | [OII], [NeIII], $H_\beta$, [OIII], [OIII], $H_\alpha$, [SII], |
| HS0752+5603 | $H_\gamma$, $H_\beta$, [OIII], [OIII], HeI, [OI], $H_\alpha$+[NII], [SII], [AIII] |
| HS0752+6147 | [OII], $H_\gamma$, $H_\beta$, [OIII], [OIII], HeI |
| HS0757+6441 | [OII], $H_\beta$, [OIII], [OIII], $H_\alpha$, [SII] |
| HS0805+5742 | $H_\gamma$, $H_\beta$, [OIII], [OIII], HeI, [OI], $H_\alpha$+[NII], [SII] |
| HS0814+6439 | Sy1: [OII], $H_\delta$, $H_\gamma$, $H_\beta$, [OIII], [OIII], $H_\alpha$, [SII], |
| HS0831+6215 | [NeIII], $H_\delta$, $H_\gamma$, [OIII], $H_\beta$, [OIII], [OIII], HeI, [OI], $H_\alpha$+[NII], HeI, [SII], [AIII] |
| HS0838+6253 | [OII], $H_\gamma$, $H_\beta$, [OIII], [OIII] -incomplete spectrum |
| HS0915+5540 | [NeIII], $H_\delta$, $H_\gamma$, [OIII], $H_\beta$, [OIII], [OIII], HeII, HeI, [OI], [OI], $H_\alpha$+[NII], [SII], [AIII] |
| HS1222+3741 | [OII], [NeIII], $H_\gamma$, $H_\beta$, [OIII], [OIII], HeI, $H_\alpha$, [SII] |
| HS1223+3938 | [OII], $H_\beta$, [OIII], [OIII], $H_\alpha$+[NII], [SII] |
| HS1232+3846 | [OII], $H_\beta$, [OIII], [OIII], $H_\alpha$+[NII], [SII] |
| HS1232+3947 | [NeIII], $H_\delta$, $H_\gamma$, $H_\beta$, [OIII], [OIII], HeI, $H_\alpha$, [SII] |
| HS1232+3612 | [OII], [NeIII], $H_\gamma$, $H_\beta$, [OIII], [OIII], HeI, [OI], $H_\alpha$+[NII], [SII], [AIII] |
| HS1236+3821 | $H_\beta$, [OIII], [OIII], $H_\alpha$+[NII], [SII] |
| HS1244+3648 | [OII], $H_\beta$, [OIII], [OIII], $H_\alpha$+[NII], [SII] |
| HS1254+3323 | abs: CaK, $G_{band}$, NaD, Mgb? |
| HS1255+3506 | [OII], $H_\beta$, [OIII], [OIII], $H_\alpha$, [SII], |
| HS1256+3505 | [OII], $H_\beta$, [OIII], [OIII], $H_\alpha$, [SII], |
| HS1256+3512 | [OII], $H_\beta$, [OIII], [OIII], $H_\alpha$, [SII], |
| HS1301+3312 | $H_\beta$, [OIII], [OIII], $H_\alpha$, [SII] |
| HS1301+3325 | [OII], $H_\beta$, [OIII], [OIII], $H_\alpha$ |
| HS1301+3209 | $H_\beta$, [OIII], [OIII], $H_\alpha$+[NII], [SII] [AIII], [OII] |
| HS1302+3046 | $H_\alpha$ |
| HS1304+3529 | $H_\gamma$, $H_\beta$, [OIII], [OIII], $H_\alpha$+[NII], [SII] |
| HS1306+3525 | [OIII], $H_\alpha$ |
| HS1306+3320 | $H_\beta$, [OIII], [OIII], $H_\alpha$, [SII] |
| HS1306+3527 | $H_\alpha$, [SII] |
| HS1308+3044 | $H_\beta$, [OIII], [OIII], $H_\alpha$+[NII], [SII] |
| HS1309+3409 | $H_\beta$, $H_\alpha$ |
| HS1309+3431 | [OIII], $H_\alpha$+[NII], [SII] |
| HS1311+3628 | [NeIII], $H_\zeta$, $H_\epsilon$, $H_\delta$, $H_\gamma$, [OIII], $H_\beta$, [OIII], [OIII], HeI, [OI], [OI], $H_\alpha$, HeI, [SII], [AIII], [OII] |
| HS1312+3847 | [OII], [NeIII], $H_\zeta$, $H_\epsilon$, $H_\delta$, $H_\gamma$, $H_\beta$, [OIII], [OIII], HeI, [OI], $H_\alpha$,[SII], [AIII] |
| HS1312+3508 | $H_\gamma$, $H_\beta$, [OIII], [OIII], HeI, [OI]+[SII], [OI], $H_\alpha$ HeI, [SII], HeI, [AIII], [OII], [AIII] |
| HS1318+3406 | $H_\beta$, [OIII], [OIII], $H_\alpha$, [SII] |
| HS1323+3211 | $H_\alpha$ |
| HS1323+3319 | $H_\gamma$, $H_\beta$, [OIII], [OIII], HeI, [OI], $H_\alpha$, [SII] |
| HS1328+3424 | $H_\beta$, [OIII], [OIII], $H_\alpha$+[NII], [SII] |
| HS1329+3703 | [OII], $H_\beta$, [OIII], [OIII], $H_\alpha$+[NII], [SII] |
| HS1330+3651 | [NeIII], $H_\epsilon$, $H_\delta$, $H_\gamma$, $H_\beta$, [OIII], [OIII], HeI, [OI], $H_\alpha$+[NII], [SII], [AIII] |
| HS1331+3906 | $H_\alpha$+[NII], [SII] |
| HS1333+3149 | [OIII], $H_\alpha$+[NII] |

**Table 5: Continued**

| object name | emission lines |
|---|---|
| HS1333+3058 | $H_\gamma$, $H_\beta$, [OIII], [OIII], HeI, $H_\alpha$, [SII] |
| HS1334+3957 | [NeIII], $H_\epsilon$, $H_\delta$, $H_\gamma$, [OIII], $H_\beta$, [OIII], [OIII], HeI, $H_\alpha$+[NII], [SII], [AIII] |
| HS1336+3114 | $H_\beta$, [OIII], [OIII], $H_\alpha$+[NII], [SII] |
| HS1336+3650 | $H_\beta$, [OIII], [OIII], $H_\alpha$ |
| HS1340+3307 | $H_\beta$, [OIII], [OIII], HeI, $H_\alpha$, [SII] |
| HS1340+3207 | $H_\beta$, [OIII], [OIII], $H_\alpha$, [SII] |
| HS1341+3409 | $H_\beta$, [OIII], [OIII], $H_\alpha$, [SII] |
| HS1349+3942 | $H_\beta$, [OIII], [OIII], $H_\alpha$+[NII], [SII] |
| HS1354+3634 | $H_\gamma$, $H_\beta$, [OIII], [OIII], HeI, [OI], $H_\alpha$+[NII], [SII] |
| HS1354+3635 | $H_\beta$, [OIII], [OIII], $H_\alpha$+[NII], [SII] |
| HS1402+3650 | [OII], $H_\gamma$, $H_\beta$, [OIII], [OIII], HeI, $H_\alpha$+[NII], [SII] |
| HS1408+4201 | abs: Mgb, NaD |
| HS1410+3627 | [OII], $H_\beta$, [OIII], [OIII], $H_\alpha$, [SII] |
| HS1413+4402 | [OII], $H_\beta$, [OIII], [OIII], $H_\alpha$+[NII], [SII] |
| HS1415+4203 | $H_\alpha$+[NII] |
| HS1416+3554 | [NeIII], $H_\beta$, [OIII], [OIII], HeI, [OI], $H_\alpha$+[NII], [SII] |
| HS1420+3437 | $H_\beta$, [OIII], [OIII], HeI, [OI], $H_\alpha$, [SII] |
| HS1421+4018 | abs: CaH, $G_{band}$, Mgb, NaD |
| | emi: [OII], $H_\alpha$+[NII]?, [SII] |
| HS1422+3325 | $H_\beta$, [OIII],[OIII], HeI, [OI], $H_\alpha$, [SII] |
| HS1422+3339 | $H_\beta$, [OIII], [OIII], $H_\alpha$, [SII] |
| HS1425+3835 | $H_\beta$, [OIII], [OIII], $H_\alpha$+[NII], [SII] |
| HS1429+3451 | $H_\gamma$, $H_\beta$, [OIII], [OIII], HeI, [OI], $H_\alpha$, HeI?, [SII] |
| HS1429+3154 | $H_\gamma$, $H_\beta$, [OIII], [OIII], HeI, $H_\alpha$, [SII] |
| HS1429+4511 | $H_\beta$, [OIII], [OIII], $H_\alpha$+[NII], [SII] |
| HS1435+4523 | [OII], $H_\beta$, [OIII], [OIII], $H_\alpha$+[NII], [SII] |
| HS1438+3147 | [OII], [NeIII], $H_\gamma$, [OIII], $H_\beta$, [OIII], [OIII], HeI, [OI], $H_\alpha$, [SII] |
| HS1440+3805 | $H_\beta$, [OIII], [OIII], HeI, [OI], $H_\alpha$+[NII], [SII] |
| HS1440+4302 | [NeIII], $H_\gamma$, $H_\beta$, [OIII], [OIII], HeI, [OI], $H_\alpha$+[NII], [SII], [AIII] |
| HS1442+4250 | [NeIII], $H_\zeta$, $H_\epsilon$, $H_\delta$, $H_\gamma$, [OIII], $H_\beta$, [OIII], [OIII], HeI, [OI], $H_\alpha$, [SII], [AIII] |
| HS1442+4332 | $H_\alpha$ ? |
| HS1444+3114 | $H_\beta$, [OIII], [OIII], $H_\alpha$+[NII], [SII] |
| HS1502+4152 | $H_\beta$, [OIII], [OIII], $H_\alpha$, [SII] |
| HS1505+3944 | $H_\alpha$, [SII] |
| HS1507+3743 | [OII], [NeIII], $H_\zeta$, $H_\epsilon$, $H_\delta$, $H_\gamma$, [OIII], $H_\beta$, [OIII], [OIII], HeI, [OI], $H_\alpha$, [SII], [AIII] |
| HS1522+4214 | $H_\beta$, [OIII], [OIII], $H_\alpha$+[NII], [SII] |
| HS1524+4205 | [OIII], [OIII], $H_\alpha$+[NII], [SII] |
| HS1526+4045 | [OIII], $H_\alpha$, [SII] |
| HS1543+4525 | $H_\beta$, [OIII], [OIII], HeI, [OI], $H_{alpha}$, [SII], [AIII] |
| | emi: [OII], [SII] |
| HS1544+4736 | $H_\gamma$, $H_\beta$, [OIII], [OIII], HeI, $H_\alpha$, [SII], [AIII] |
| HS1546+4755 | $H_\beta$, [OIII], [OIII], $H_\alpha$, [SII] |
| HS1609+4827 | $H_\beta$, [OIII], [OIII], HeI?, [OI]?, $H_\alpha$+[NII], [SII] |
| HS1610+4539 | $H_\gamma$, $H_\beta$, [OIII], [OIII], HeI, [OI], $H_\alpha$, [SII], [AIII] |
| HS1614+4709 | $H_\gamma$, $H_\beta$, [OIII], [OIII], HeI, [OI], $H_\alpha$, [SII], [AIII] |
| HS1633+4703 | $H_\beta$, [OIII], [OIII], $H_\alpha$, [SII] |
| HS1640+5136 | $H_\gamma$?, HeII?, $H_\beta$, [OIII], [OIII], HeI, [OI], $H_\alpha$+[NII], [SII] |
| HS1641+5053 | $H_\beta$, [OIII], [OIII], HeI, [OI], $H_\alpha$+[NII], [SII] |
| HS1643+5313 | QSO: MgII, FeII |
| HS1645+5155 | $H_\beta$, [OIII], [OIII], $H_\alpha$, [SII] |
| HS1657+5033 | $H_\beta$, [OIII], [OIII], HeI, [OI], $H_\alpha$, [SII] |
| HS1657+5735 | $H_\beta$, [OIII], [OIII], HeI, [OI], $H_\alpha$, [SII] |
| HS1627+5239 | $H_\beta$, [OIII], [OIII], HeI, $H_\alpha$, [SII] |

| object name | emission lines |
|---|---|
| HS1645+5155 | $H_\beta$, [OIII], [OIII], $H_\alpha$+[NII], [SII] |
| HS1711+5758 | QSO: $Ly_\alpha$, [OI]?, CIV |
| HS1723+5631 | $H_\beta$, [OIII], [OIII], HeI, $H_\alpha$, [SII] |
| HS1728+5655 | $H_\gamma$, $H_\beta$, [OIII], [OIII], [HeI], [OI]?, $H_\alpha$, |
| HS1734+5704 | $H_\gamma$, $H_\beta$, [OIII], [OIII], HeI, [OI], $H_\alpha$, [SII] |
| HS2353+2005 | $H_\beta$, [OIII], [OIII], $H_\alpha$+[NII], [SII] |

The complete sequence of emission lines listed in the table (wavelengths given in Å): $Ly_\alpha$ 1216, NV 1240, CIV 1549, MgII 2798, [OII] (blend) 3727.45, [NeIII] 3868.76, $H_\zeta$ (blend HeI) 3889.05 (3889.65), $H_\epsilon$ (blend [NeIII], [OII]) 3970.07 (3967.47, 3967.40), $H_\delta$ 4101.6, $H_\gamma$ 4340.3, [OIII] 4363.21, HeII 4686, $H_\beta$ 4861.2, [OIII] 4958.92, [OIII] 5006.85, HeI 5875.99, [OI] (blend with [SII]) 6300.32 (6312.1), [OI] 6363.81, [NII] 6548.2, $H_\alpha$ 6562.9, [NII] 6583.6, HeI 6678.1, [SII] (blend) 6723.6, HeI 7065.3, [AIII] 7135.8, [OII] (blend) 7319.9, 7330.2, [AIII] 7751.02

described by Maddox et al. (1990), Irwin, Maddox & McMahon (1994). All objects with new redshifts have been searched in the APM catalogue facility and 146 have been retrieved by comparison of their position only. It is beyond the scope of this paper to clarify why some sources were missed. We note that there is no significant deviation between the RA and $\delta$ values from our measurements and those of the APM:

$$[RA_{(this\ paper)} - RA_{APM}] = -0.04 \pm 0.11\, sec$$
$$[\delta_{(this\ paper)} - \delta_{APM}] = -0.62 \pm 1.25\, arcsec$$

and, that the error distribution corresponds to our expectations (see above). The comparison of the B magnitudes, however, shows a surprising distribution (Fig. 6). The difference between the APM magnitudes and ours seems to increase monotonically with brightness, with a scatter of about ± 1.4 mag. superimposed. For fainter (B $\geq$ 16) galaxies one may understand the diagram in the sense of a small offset and a magnitude error in both samples in the order of 0.4 to 0.5 mag. The differences for the brighter objects however cannot be explained as before. According to the APM magnitudes, one would expect that we have NGC and even Shapley-Ames galaxies in our sample, which is not the case. The distribution ranges up to the unrealistic $B_{APM}$=11 mag. This may mean that there are still some uncertainties in the calibration of the APM magnitudes for the bright objects, where the plates tend to be saturated over a large extent of the galaxy (see also Metcalfe et al. 1995). We finally conclude that the subsample of our objects with 16 $\leq B \leq$ 19.5 have good total luminosity estimates, while for the brighter ones, our values are only lower limits. It is customary to give a magnitude for which a survey is complete, but for our sample the incompleteness are also described in terms of colour, line fluxes and equivalent widths of the emission lines. (A detailed description of the incompleteness will be given in a future paper).

If we exclude the emission-line galaxies that were found as second priority objects, we find that our objects have redshifts between $0 < z < 0.1$, which is exactly the range for which the [OIII] line and/or $H_\beta$ can be seen in the IIIa-J plates, due to the cutoff of the emulsion at 5400 Å. In Fig. 7 we give the redshift distribution of our sample. With dotted line we plot the whole sample, included the blue candidates, while with solid line we plot only the first priority objects. The histogram has a peak at z=0.015, which shows that our sample is dominated by nearby galaxies (as intended for a study of the nearby large scale structure). The histogram drops very fast after 0.06, containing only a few objects with $0.06 < z < 0.10$, most of them being blue objects. The few galaxies with $z > 0.1$ are also from the second category objects.

## 5. Conclusions

Our survey for emission-line galaxies contains a complete sample based on candidates selected from digitized objective prism plates. We used the IIIa-J plates taken in the frame of the Hamburg QSO Survey. Automated search software is applied to the digitized data to select spectra in a certain parameter space. The two selection criteria are the "brightness" and the "colour" of the spectra. The selected candidates are then rescanned with high resolution and the final digitized spectra are visually inspected for emission lines. Some second priority candidates were also selected because of their blue spectra. We observed all the first priority candidates and also some of the second priority ones.

1. The final sample contains 203 objects, of which 196 are ELG, four are galaxies with absorption and three are QSOs. Almost half of our sample contains newly discovered objects, and three quarters of the given redshifts are new.

2. The mean surface density of our sample of emission-line galaxies is 0.16. This value is lower than the value obtained by the Michigan Survey, for example, but our sample is selected with an automated procedure, in a certain interval of parameters.

3. Our galaxies have apparent magnitudes between $15.0 < B < 19.5$. A comparison with the magnitudes available in the literature, as well as with the APM ones, shows that our magnitudes are reliable only for the compact faint objects. In the case of extended objects, with B$\leq$ 16, our survey is sensitive only for the emission core, the total magnitudes being thus systematically fainter.

**Table 6: List of identified stars**

| object name | coord. | B | type | abs. lines |
|---|---|---|---|---|
| HS0007+0520 | 00 07 16.1 +05 20 42 | 19.4 | M | |
| HS0014+0351 | 00 14 34.5 +03 51 43 | 18.0 | | Mgb? |
| HS0015+1123 | 00 15 18.3 +11 23 00 | 18.4 | M | |
| HS0018+2040 | 00 18 35.1 +20 40 36 | 18.7 | G | $H_\beta$, Mgb |
| HS0019+0006 | 00 19 20.5 +00 06 30 | 17.4 | | $H_\gamma$, $H_\beta$, $H_\alpha$ |
| HS0024+2413 | 00 24 55.6 +24 13 43 | 17.3 | G | CaK, CaH, $G_{band}$, $H_\beta$, Mgb, NaI, $H_\alpha$ |
| HS0028+1201 | 00 28 31.3 +12 01 18 | 17.9 | F or G | Mgb, NaI, $H_\alpha$ |
| HS0028+2419 | 00 28 41.7 +24 19 50 | 17.8 | G | $G_{band}$, Mgb, $H_\alpha$ |
| HS0030+1433 | 00 30 17.2 +14 33 09 | 17.4 | F | $G_{band}$, $H_\beta$, Mgb, $H_\alpha$ |
| HS0039+2151 | 00 39 53.0 +21 51 37 | 16.8 | F | $G_{band}$, $H_\alpha$ |
| HS0040+1344 | 00 40 28.8 +13 44 15 | 18.2 | F or G | $H_\beta$, NaD, $H_\alpha$ |
| HS0103+1520 | 01 03 19.3 +15 20 20 | 17.6 | late-type star | Mgb, $H_\alpha$ |
| HS0107+0006 | 01 07 58.7 +00 06 53 | 17.8 | G | CaK, CaH, $H_\beta$, $H_\alpha$ |
| HS0109+1540 | 01 09 30.0 +15 40 07 | 15.8 | early-type star | $H_\beta$, $H_\alpha$ |
| HS0739+6019 | 07 39 07.4 +60 19 32 | 18.0 | G | $G_{band}$, $H_\beta$, Mgb, $H_\alpha$ |
| HS0750+6337 | 07 50 44.4 +63 37 29 | 18.1 | M | |
| HS0753+5645 | 07 53 41.8 +56 45 12 | 17.1 | G | CaK, CaH, $G_{band}$, $H_\beta$, $H_\alpha$ |
| HS0815+5658 | 08 15 58.9 +56 58 19 | 17.4 | late-type star | Mgb, NaD, $H_\alpha$ |
| HS1311+3350 | 13 11 21.4 +33 50 01 | 17.7 | A | CaH, $H_\delta$, $H_\gamma$, $H_\beta$, $H_\alpha$ |
| HS1313+3127 | 13 13 29.9 +31 27 04 | 18.0 | | Mgb, $H_\alpha$ |
| HS1315+3340 | 13 15 17.3 +33 40 27 | 18.4 | | $H_\beta$, NaI |
| HS1328+3528 | 13 28 28.4 +35 28 39 | | F or G | Cak, CaH, $G_{band}$, $H_\beta$, $H_\alpha$ |
| HS1406+3427 | 14 06 42.5 +34 27 55 | 16.5 | G | $G_{band}$, FeI, $H_\beta$, Mgb, NaD, $H_\alpha$ |
| HS1406+3844 | 14 06 22.2 +38 44 39 | | A or F | $H_\delta$, $G_{band}$, $H_\gamma$, $H_\beta$, $H_\alpha$ |
| HS1416+3932 | 14 16 30.3 +39 32 11 | | F | CaK, CaH, $G_{band}$, $H_\beta$, $H_\alpha$ |
| HS1419+4116 | 14 19 29.9 +41 16 14 | | M | |
| HS1427+4200 | 14 27 12.9 +42 00 58 | | A or F | CaK, CaH, $H_\beta$, $H_\alpha$ |
| HS1432+4255 | 14 32 38.5 +42 55 30 | | M | |
| HS1445+4412 | 14 45 17.9 +44 12 06 | | | $H_\alpha$ |
| HS1448+3952 | 14 48 32.3 +39 52 39 | | F or G | CaK, $G_{band}$, $H_\beta$, $H_\alpha$ |
| HS1452+3009 | 14 52 25.2 +30 09 06 | | A | CaK, CaH, $H_\delta$, $G_{band}$, $H_\beta$, $H_\alpha$ |
| HS1502+3444 | 15 02 38.7 +34 44 55 | | F or G | CaK, CaH, $G_{band}$, $H_\beta$, Mgb, NaD, $H_\alpha$ |
| HS1503+3822 | 15 03 21.8 +38 22 41 | | A or F | CaK, CaH, $H_\delta$, $H_\gamma$, $H_\beta$, $H_\alpha$ |
| HS1508+4025 | 15 08 05.2 +40 25 29 | | F | CaK, CaH, $H_\delta$, $H_\gamma$, $H_\beta$, Mgb, $H_\alpha$ |
| HS1510+3551 | 15 10 40.6 +35 51 16 | | late-type star | CaK, CaH |
| HS1516+4441 | 15 16 19.1 +44 41 45 | | M | |
| HS1517+4021 | 15 17 55.0 +40 21 32 | | F or G | CaK, CaH, $G_{band}$, $H_\beta$, Mgb, $H_\alpha$ |
| HS1523+4420 | 15 23 22.3 +44 20 38 | | F | $H_\gamma$, $H_\beta$, Mgb, $H_\alpha$ |
| HS1547+4708 | 15 47 14.2 +47 08 45 | 18.4 | | $H_\beta$, $H_\alpha$ |
| HS1548+4555 | 15 48 07.4 +45 55 20 | 18.5 | | Mgb, $H_\alpha$ |
| HS1611+4825 | 16 11 57.9 +48 25 35 | 18.1 | A or F | $H_\gamma$, $H_\beta$, $H_\alpha$ |
| HS1612+4505 | 16 12 45.5 +45 05 57 | 17.5 | F? | $H_\beta$, Mgb, $H_\alpha$ |
| HS1626+5132 | 16 26 47.2 +51 32 05 | 18.0 | A or F | $H_\beta$, $H_\alpha$ |
| HS1639+5103 | 16 39 43.1 +51 03 43 | 18.5 | A | $H_\gamma$, $H_\alpha$ |
| HS1655+5209 | 16 55 56.6 +52 09 42 | 17.3 | F or G | $G_{band}$, $H_\gamma$, $H_\beta$, $H_\alpha$ |
| HS1657+5207 | 16 57 47.0 +52 07 33 | 19.5 | F or G | $H_\beta$, Mgb, $H_\alpha$ |
| HS1721+5819 | 17 21 23.1 +58 19 51 | 16.1 | G | $G_{band}$, $H_\gamma$, FeI?, $H_\beta$, Mgb |

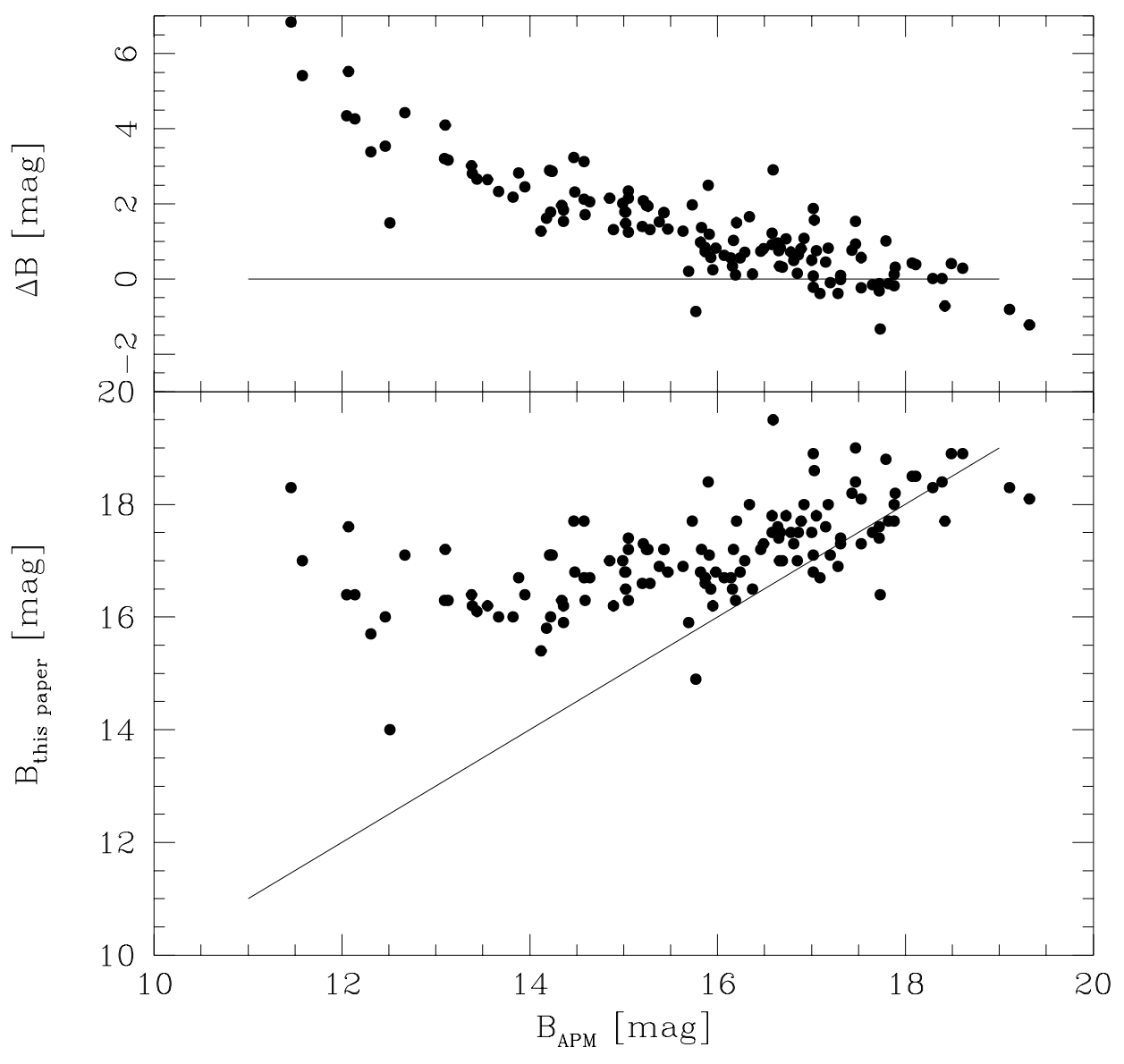

**Fig. 6.** The same as in Fig. 5 but for the comparison with the APM magnitudes.

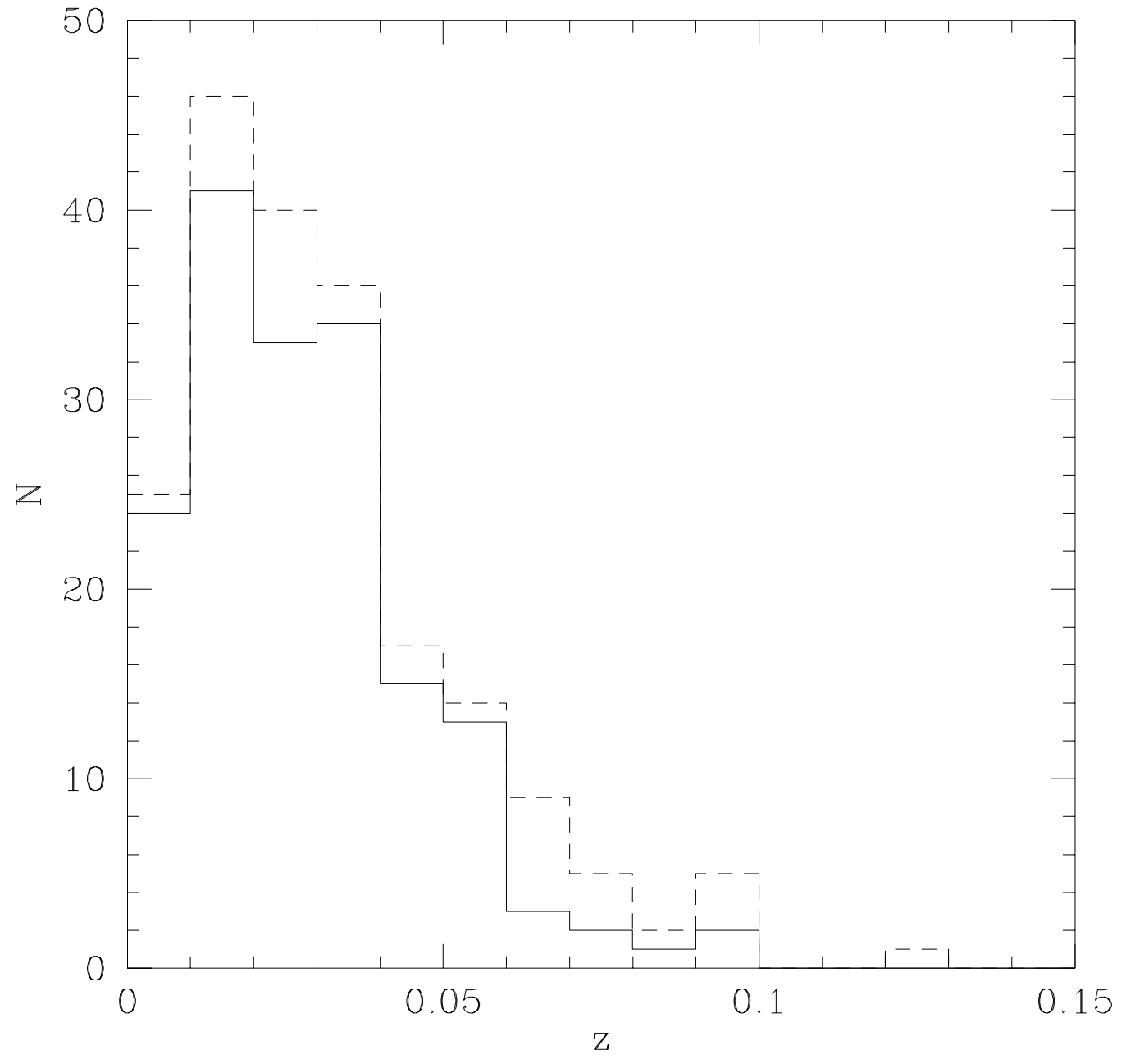

**Fig. 7.** The redshift distribution of our survey. With solid lines only the first priority objects are plotted, while with dotted lines all objects are included, also the second priority objects.

4. The redshift distribution ranges between $0 < z < 0.1$, with a peak at z=0.015, which show that we found a sample of relatively nearby galaxies. The few galaxies with $z > 0.1$ were either selected as blue objects, or are Sy 1 galaxies.

*Acknowledgements.* We would like to thank Dr. A.P. Fairall for the careful review of this manuscript and Dr. Bernd Kuhn for observing some of our objects during his run in La Silla, December, 1994, as well as for his support with the reduction software. We gratefully acknowledge Dr. Dieter Engels for processing the magnitude calibration of the Hamburg prism plates. C. C. Popescu is greatly indebted to the astronomers from Hamburger Sternwarte for their warm hospitality during the work with their Plate Archive. It is a pleasure to thank the Calar Alto staff for their support during observations and Dr. Klaus Meisenheimer for his support during the Cafos22 run. U. Hopp acknowledge the support by Sonderforschungsbereich 375 of the Deutsche Forschungsgemeinschaft during the end of this project.

This research has made use of the NASA/IPAC Extragalactic Database (NED) which is operated by the Jet Propulsion Laboratory, California Institute of Technology, under contract with the National Aeronautics and Space Administration.

## References

Acker, A., Ochsenbein, F., Stenholm, B., Tylenda, R., Marcout, J., Schohn, C. 1992, The Strasbourg-ESO Catalogue of Galactic Planetary Nebulae, published by the European Southern Observatory

Alonso, O., Zamorano, J., Rego, M. and Gallego, J. 1995, A&AS 113, 1

Bardeen, J.M. 1986, in *Inner Space/Outer Space*, ed. E.W. Kolb, M.S. Turner, D. Lindley, K. Olive, and D. Seckel (University of Chicago Press, Chicago), p.212

Binggeli, B., Tarenghi, M., Sandage, A. 1990, A&A 228, 42

Bohuski, T.J., Fairall, A.P., and Weedman, D.W. 1978, ApJ 221, 776

Clowes, R.G., Cooke, J.A., Beard, S.M. 1984, MNRAS 207, 99

Cooke J.A., Beard, S.M., Emerson, D., Kelly, B.D., MacGillivray, H.T. 1986, MNRAS 219, 241

Coziol, R., Demers, S., Peña, M., Torres-Peimberg, S., Fontaine, G., Wesemael, F., Lamontagne, R. 1993, AJ 105(1), 35 (Paper I)

Coziol, R., Demers, S., Peña, M., Barnéoud, R. 1994, AJ 108, 405 (Paper II)

Engels, D., Cordis, L., Köhler, T. 1994, in IAU Symposium 161, ed. H.T. MacGillivray, Kluwer, Dordrecht, p. 317

Gallego, J. 1995, PhD thesis (Madrid)

Hagen, H.-J., Groote, D., Engels, D., Reimers, D. 1995, A&AS 111, 195

Haro, G. 1956, Bol. Obs. Tonantzintla y Tacubaya 14, 329

Hewett, P.C., Irwin, M.J., Bunclark, P. et al. 1985, MNRAS 213, 971

Horne, K. 1986, PASP 98, 609

Hopp, U., and Kuhn, B. 1995, in Astron. Ges. Review Ser. 8, p. 277, ed. G. Klare

Hopp, U., Kuhn, B., Thiele, U., Birkle, K., and Elsässer H. 1995, A&AS 109, 537

Irwin, M., Maddox, S., and McMahon, R. 1994, RGO Spectrum 2, 14

Irwin, M.J., and Trimble, V. 1984, AJ 89, 83

Kaiser, N. 1986, in *Inner/Space/Outer Space*, see ref. Bardeen (1986), p.258

Kibblewhite E.J., Bridgeland, M.T., Bunclark, P., Irwin, M. 1984, in: Astronomical Microdensitometry Conference, ed. D.A. Klinglesmith, NASA Conf. Pub. 2317, p. 277

Kinman, T.D. 1984, in Astronomy with Schmidt-Type Telescopes, ed. M. Capaccioli (Dordrecht: Reidel), p. 409

Kunth, D., Sargent, W.L.W., and Kowal, C. 1981, A&AS 44, 229

Kunth, D. and Sargent, W.L.W. 1986, AJ 91, 761

Maddox, S.J., Efstathiou, G., Sutherland, W. and Loveday, J. 1990, MNRAS 242, 43

Markarian, B.E. 1967, Astrofizika 3, 55

Markarian, B.E., Lipovetskii, V.A., and Stepanian, D.A. 1983a, Astrofizika 19, 29; Astrophysics 19, 14

Markarian, B.E., Lipovetskii, V.A., and Stepanian, D.A. 1983b, Astrofizika 19, 221

Markarian, B.E., Lipovetskii, V.A., and Stepanian, D.A. 1984, Astrofizika 20, 419

Maza, J., Ruiz, M.T., Gonzalez, L.E., and Wischnjewsky, M. 1989, ApJS, 69 (list 1)

MacAlpine, G.M., Smith, S.B., and Lewis, D.W. 1977a, ApJS 34, 95 (List I)

MacAlpine, G.M., Smith S.B., and Lewis, D.W. 1977b, ApJS 35, 197 (List II)

MacAlpine, G.M., Lewis, D.W., Smith, S.B. 1977c, ApJS 35, 203 (List III)

MacAlpine, G.M., Lewis, D.W. 1978, ApJS 36, 587 (List IV)

MacAlpine G.M. and Williams, G.A. 1981, ApJS 45, 113 (List V)

Metcalfe, N., Fong, R., Shanks, T. 1995, MNRAS 274, 679

Moody, J.W., Kirshner, R.P., MacAlpine, G.M., and Gregory, S.A. 1987, ApJ 314, L33

Pesch, P., Sanduleak, N. (Paper I) 1983, ApJS 51, 171

Pesch, P., Sanduleak, N. (Paper III) 1986, ApJS 60, 543

Pesch, P., Sanduleak, N. (Paper V) 1988, ApJS 66, 297

Pesch, P., Sanduleak, N. (Paper VIII) 1989, ApJS 70, 163

Pesch, P., Sanduleak, N., and Stephenson, C.B. (Paper XII) 1991, ApJS 76, 1043

Pesch, P., Stephenson, C.B., and MacConnell D.J. 1995, ApJS 98, 41

Salzer J.J. 1989, ApJ 347, 152

Salzer, J., MacAlpine, G.M., and Boroson, T.A. 1989, ApJS 70,447

Salzer, J., Moody, J.W., Rosenberg, J.L., Gregory, S.A., and Newberry, M.V. 1995, AJ 109, 2376

Sanduleak, N. and Pesch, P. 1982, ApJ 258, L11

Sanduleak, N., Pesch, P. (Paper II) 1984, ApJS 55, 517

Sanduleak, N., and Pesch, P. (Paper IV) 1987, ApJS 63, 809

Sanduleak, N., Pesch, P. (Paper IX) 1989, ApJS 70, 173

Sanduleak, N., Pesch, P. (Paper XI) 1990, ApJS 72, 291

Schneider, D.P., Schmidt, M., Gunn, J.E. 1994, AJ 107, 1245

Smith, M.G. 1975, ApJ 202, 591

Smith, M.G., Aguirre, C. and Zemelman, M. 1976, ApJS 32, 217 (List N1)

Stepanian, J.A., Lipovetsky, V.A., and Erastova, L.K. 1990, Astrofizika 32, 441; Astrophysics 32, 252

Stepanian, J.A., Lipovetsky, V.A., and Erastova, L.K. 1991, Astrofizika 34, 205; Astrophysics 34, 99

Stephenson, B., Pesch, P. (Paper XII) 1992, ApJS 82, 471

Stickel, M., Fried, J.W., Kúhr, H. 1993, A&AS 98, 393

Surace, C. 1993, PhD thesis (Marseille)

Surace, C., Comte, G. 1994, A&A 281(3), 653

Takase, B., and Miyauchi-Isobe, N. 1988, Ann. Tokyo Astr. Obs., Ser. 2, 22, 41

Thuan, T.X., Gott, J.R., Schneider, S.E. 1991, ApJ 315, L93

Wamsteker, W., Prieto, A., Vitores, A., Schuster, H.E., Danks, A.C., Gonzalez, R., and Rodriguez, G. 1985, A&AS 62, 255

Wasilewski, A.J. 1983, ApJ 272, 68

Weistrop, D., Hintzen, P., Kennicutt, R.C., Liu. C., Lowenthal, J., Cheng, K.-P., Oliversen, R., Woodgate, B. 1992, ApJ 396, L23

Wisotzki, L. 1994, in IAU Symposium 161, eds. H.T. MacGillivray et. al. (Kluwer, Dordrecht) 723

Zamorano, J., Rego, M., Gonzáles-Riestra, R., and Rodrígues, G. 1990, Ap&SS 170, 353

Zamorano, J., Rego, M., Gallego, J., Vitores, A.G., Gonzáles-Riestra, R., and Rodrígues-Caderot, G. 1994, ApJS 95, 387

Zeldovich, Y.B., Einasto, J., and Shandarin, S.F. 1982, Nature 300, 407